\documentclass[aip]{revtex4}
\usepackage{graphicx}
\usepackage{dcolumn}
\usepackage{amssymb,amsmath}
\usepackage{natbib}
\usepackage{subfigure}
\usepackage{epstopdf}
\usepackage{placeins}
\usepackage{color}
\usepackage{epstopdf}

\begin{document}

The following article has been accepted by the Journal of Renewable and Sustainable Energy. After it is published, it will be found at http://scitation.aip.org/content/aip/journal/jrse.

\title{Coupled wake boundary layer model of wind-farms}
\author{Richard J. A. M. Stevens$^{1,2}$ Dennice F. Gayme$^{1}$ and Charles Meneveau}
\affiliation{
$^1$Department of Mechanical Engineering $\&$ Center for Environmental and Applied Fluid Mechanics, Johns Hopkins University, Baltimore, Maryland 21218, USA\\
$^2$Department of Physics, Mesa+ Institute, and J.\ M.\ Burgers Centre for Fluid Dynamics, University of Twente, 7500 AE Enschede, The Netherlands}

\date{\today}

\begin{abstract}
We present and test the coupled wake boundary layer (CWBL) model that describes the distribution of the power output in a wind-farm. The model couples the traditional, industry-standard wake model approach with a ``top-down'' model for the overall wind-farm boundary layer structure. This wake model captures the effect of turbine positioning, while the ``top-down'' portion of the model adds the interactions between the wind-turbine wakes and the atmospheric boundary layer. Each portion of the model requires specification of a parameter that is not known a-priori. For the wake model, the wake expansion coefficient is required, while the ``top-down'' model requires an effective spanwise turbine spacing within which the model's momentum balance is relevant. The wake expansion coefficient is obtained by matching the predicted mean velocity at the turbine from both approaches, while the effective spanwise turbine spacing depends on turbine positioning and thus can be determined from the wake model. Coupling of the constitutive components of the CWBL model is achieved by iterating these parameters until convergence is reached. We illustrate the performance of the model by applying it to both developing wind-farms including entrance effects and to fully developed (deep-array) conditions. Comparisons of the CWBL model predictions with results from a suite of large eddy simulations (LES) shows that the model closely represents the results obtained in these high-fidelity numerical simulations. A comparison with measured power degradation at the Horns Rev and Nysted wind-farms shows that the model can also be successfully applied to real wind-farms. \\
\end{abstract}


\maketitle

\section{Introduction} \label{Section_Introduction}
It is well known that wakes created by upstream wind-turbines can significantly influence the power production of downstream turbines in wind-farms \cite{bar09b,bar09c,ste14b}. Modeling wake effects is important in order to estimate the power production of different wind-farm layouts \cite{arc14}. Especially for large wind-farms, the two-way coupling of the relevant wake-turbine interactions dynamics to the overall structure of the atmospheric boundary layer is an important factor that affects the performance of wind-farms \cite{bar09b}. Analytical modeling of these two main aspects of the problem have traditionally relied on two quite different approaches. The first approach is based on a model of the wind-turbine wakes, in which the wake diameter is assumed to expand (typically linearly) behind the turbine and the velocity deficit is obtained assuming mass (or linearized momentum) conservation \cite{lis79,jen83,kat86,cho13,pen14b,pen14,bas14,nyg14}. This procedure can be considered a ``bottom-up'' approach, which is built into typical commercial packages that are used to predict wind-farm performance. When many wakes are superposed in large wind-farms, additional complexities arise due to the vertical structure of the atmospheric boundary layer and the associated wake-atmosphere interactions are not typically captured by wake models.

\begin{figure}
\centering
\subfigure{\includegraphics[width=0.90\textwidth]{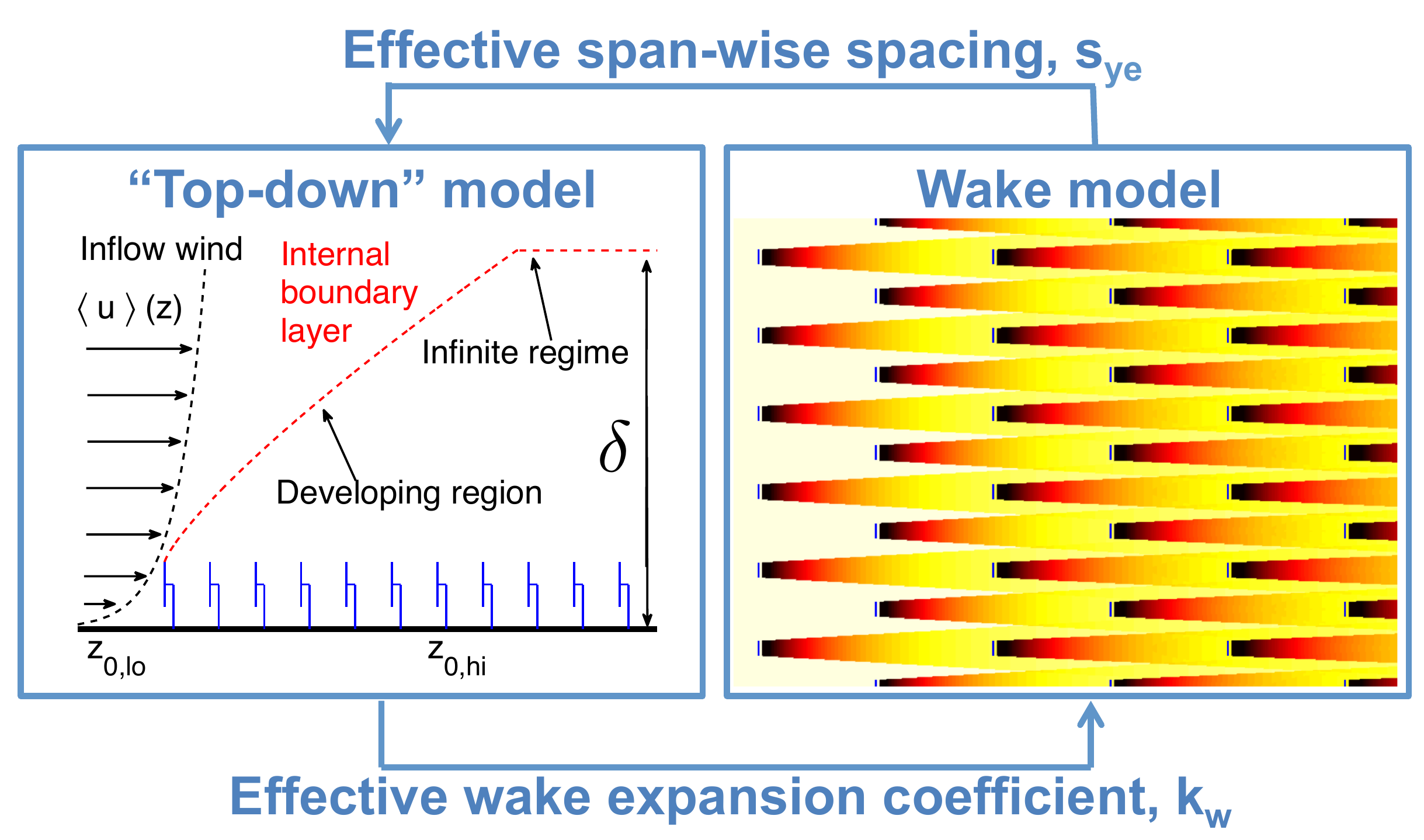}}
\caption{Conceptual sketch of the coupling between the wake and the ``top-down'' model. The ``top-down'' part captures the deep farm effects and is used, via iterations, to determine the wake expansion coefficient needed in the wake model in order to accurately capture the fully developed regime of the wind-farm. Conversely, the ``top-down'' model requires specification of an effective spanwise spacing. This distance depends on the turbine positioning and is determined with the wake model. Convergence to a consistent CWBL system is obtained by iterating until the mean streamwise velocity at turbine hub-height is the same in both models for the fully developed region of the wind-farm. The method is described in detail in \S \ref{Section_Combined}.}
\label{figure1}
\end{figure}

The second analytical approach for modeling wind-farms consists of representing the flow in an entire wind-turbine array region based on horizontal averaging. In this method, which can be considered a ``top-down'' or single-column modeling approach, the turbines are seen as roughness elements. In this framework, the average velocity profile at hub-height can be obtained based on the assumption of the existence of two logarithmic regions, one above the turbine hub-height and one below \cite{new77,jen78,fra92,eme93,fra06}. The ``top-down'' approach can predict the effective roughness height of the wind-farm. In the Calaf {\it et al.} model \cite{cal10} some wake effects are also included, although the results and predictions depend only on the area-averaged turbine spacing. Therefore, the effects based on the specific spatial arrangement of the wind-turbines, e.g.\ distinguishing between aligned and staggered configurations, see sketch in figure \ref{figure2}, is not possible. Recently this work has been extended to include predictions for the power development in large wind-farms by Meneveau \cite{men12} and Stevens \cite{ste14c}. These models have also been used to predict the optimal average turbine spacing, by taking the cost of the turbines and the land into account \cite{mey12,ste14c}. Extensions towards different atmospheric stability conditions have also been developed \cite{ses14,pen14b}.

The benefit of wake models is that they are practical and easy to implement \cite{her14}. Wake models typically perform well for predictions of the power output of turbines in the entrance region of the wind-farm, where the wake-wake interaction and the interaction with the atmosphere are limited. However, the ability of wake models to make realistic predictions degrades in the fully developed region \cite{bar09b,son14,has09,sch09,bea12,bro12} of the wind-farm. The ``top-down'' model on the other hand captures the interaction between the fully developed regime of wind-farm and the atmospheric boundary layer well, but does not include any information about the relative turbine positioning. For that reason the ``top-down'' model has difficulties predicting the turbine power output in the entrance region of the wind-farm and differences caused by the relative positioning of the turbines. Ideally, one would wish to combine both approaches and allow each to predict complementary features of the flow. To-date both the wake model and the ``top-down'' model have been applied without two-way coupling, as we will propose in the present work in an effort to combine the positive aspects of each approach.

Prior efforts at combining both approaches include the original work of Frandsen \cite{fra06}, in which three regimes are identified. In regime 1 of that model the wakes are expected to expand axisymmetrically. In regime 2 the wakes merge and specific expansion rates for the wakes are proposed. Further downstream, in regime 3, the wind-farm performance is estimated with a ``top-down'' approach like the one presented in Ref.\ \cite{fra92}. This model has led to using the ``top-down'' model as an ``upper limit'' in commercial codes \cite{has09,sch09,bea12,bro12}. Another commonly used approach is to set the wake expansion coefficient based on the turbulence intensity of the incoming flow. This was first proposed by Lissaman \cite{lis79}, and similar ideas can be found in Frandsen \cite{fra06} as well as in Yang, Kang \& Sotiropoulos \cite{yan12}. More recently, Pe\~na and Rathmann \cite{pen14b,pen13c,pen14} evaluated the effects of atmospheric stratification using the ``top-down'' infinite wind-farm boundary layer model by Frandsen \cite{fra06}. This model was extended to include atmospheric stability effect by Emeis \cite{eme10b}, to predict the wake expansion coefficient $k_\mathrm{w}$ that should be used in the Òbottom-upÓ wake model. As in Frandsen \cite{fra06} they relate $k_\mathrm{w}$ to atmospheric turbulence characteristics such as the friction velocity and turbulence intensity \cite{pen13b}.

Evaluating models based on field data from operational wind-farms is sometimes possible but it is generally very difficult due to the limited availability and lack of control over the flow parameters for the field sites. Conversely, high-fidelity numerical simulations can provide data that can be used to test simplified engineering models under idealized and well-controlled conditions. State-of-the-art Large Eddy Simulations (LES), which only require parameterizations of the smallest turbulent scales, can be utilized for this purpose. Recently, LES have been used to obtain parameterizations of the roughness height of wind-farms with an improved ``top-down'' model approach \cite{cal10,men12,ste14c}, thus describing the entire wind-farm as a roughened surface with increased momentum flux and kinetic energy extraction.

As LES requires a significant computational effort, industry still relies on less expensive methods in order to design and optimize wind-farm layouts. For example the wake model described above \cite{jen83,kat86} is used in several optimization studies \cite{mar08,ema10,kus10,saa11}. Other examples include the use of parabolized forms of the Reynolds-Averaged Navier Stokes (RANS) equations such as the Ainslie model \cite{ain88} and UPMPARK, which uses a $k-\epsilon$ turbulence model, and was later improved into WakeFarm (see e.g.\ Schepers and van der Pijl \cite{pij06,sch07b}) and Farmflow (see e.g.\ Eecen and Bot \cite{eec10}, Schepers \cite{sch12}, and \"Ozdemir {\it et al.} \cite{ozd13}), or models that are based on a parametrization of the internal boundary layer growth coupled with some eddy viscosity model, e.g.\ the Deep-Array Wake Model of Openwind \cite{bro12}. Other approaches include the Large Array wind-farm model in WindFarmer \cite{has09,sch09} and linearized CFD (computational fluid dynamics) models such as FUGA \cite{ott11}, Windmodeller \cite{bea12}, Ellipsys \cite{iva08}, and the Advanced Regional Prediction System (ARPS) \cite{xue00,xue01}. The above is not a comprehensive list. For reviews of these and additional methods we refer to Refs.\ \cite{ver03,cre99,bar09b,ret09,san11,cab11,bea12}.

In this paper we introduce the Coupled Wake Boundary Layer (CWBL) model, which provides a method of coupling the wake model \cite{jen83,kat86,pen14b} and the ``top-down'' model \cite{cal10} to provide improved predictions of the mean velocity distributions in a wind-farm and to estimate the associated wind-turbine power outputs. The wake model within the CWBL model ensures that the relative positioning of the turbines is represented, while the fully developed wind-farm's vertical structure is captured with the ``top-down'' portion of the model. Both the ``top-down'' and the wake model part of the CWBL system each contain a parameter that is not known a priori. These two parameters can be obtained from the complementary part of the CWBL model using an iterative procedure as shown schematically in figure \ref{figure1}. Here the wake growth coefficient required for the wake model is obtained by matching the predicted mean velocities or mean power with the predictions from the ``top-down'' model. Similarly the effective spanwise spacing needed by the ``top-down'' model is specified using the wake model. Being an analytical model (as opposed to differential equations based models such as RANS or LES), the CWBL model inherits the practical advantages of wake model type approaches. 
 
As an initial step, the model only considers wind-farms in which the turbines are placed on a regular lattice and the extension of this method to general geometries will be discussed in \cite{ste15}. As we are interested in developing a better understanding of the main physical mechanisms that are important for modeling and understanding the performance of very large wind-farms we have made a number of simplifications that will be justified when introduced in the following sections.

Before the coupling between both models is presented, we first briefly review the basic concepts of the wake model (section \ref{Section_Jensen}) and the ``top-down'' model (section \ref{Section_topdown}), and illustrate their previously mentioned merits and drawbacks by comparing their respective predictions with LES data. In section \ref{Section_Combined} the two-way coupling of the models is discussed in detail. This is followed by detailed comparisons of the model results with LES data, in section \ref{Section_Results}. The LES data we use are for wind-farms with $10$ or more downstream turbine rows with different combinations of spanwise and streamwise spacings. For details about the simulations we refer the reader to Refs.\ \cite{ste13,ste14b,ste14f}. In section \ref{Section_HornsRev} the model is compared to measurements from Horns Rev and Nysted. Section \ref{Section_Conclusions} provides general conclusions and an outlook to future work. 

\section{Wake model}
\label{Section_Jensen}

The classic wake model has been developed based on successive contributions by Lissaman \cite{lis79}, Jensen \cite{jen83} and Kat\'ic {\it et al.} \cite{kat86} and is also referred to as the Jensen/Park model in the literature. It was shown by Nygaard \cite{nyg14} that with a simple wake model close to that implemented in the WAsP model the power degradation data from various wind-farms (e.g.\ London Array and Nysted) could be predicted well. In addition, the author states that they find ``no robust evidence of the deep array effect''. As will be shown below, in some specific conditions the wake models indeed yield good predictions. However, they will be shown to yield incorrect predictions for very long wind-farms with staggered configurations. Also in Ref.\ \cite{nyg14}, some cases showed marked differences between data and wake models. The wake model assumes that wind-turbine wakes grow linearly (based on the notion that the background turbulence provides a spatially constant level of transverse velocity fluctuations \cite{lis79}).

In a classic far wake, conservation of linear momentum leads to the constancy of the integral of the velocity defect profile \cite{bat00}. Furthermore, in the piecewise linear profile assumed in the wake model \cite{lis79,jen83}, conservation of mass also leads to the same result \cite{bas14}. This implies that the velocity in the wake evolves according to \cite{lis79,jen83}:
\begin{equation}
u=u_{0} \left(1- \frac{1-\sqrt{1-C_\mathrm{T}}}{(1+k_\mathrm{w} x /R)^2} \right) = u_{0} \left(1- \frac{2a}{(1+k_\mathrm{w} x /R)^2} \right),
\end{equation}
where $u_{0}$ is the incoming free stream velocity, $k_\mathrm{w}$ is the wake expansion coefficient, $R$ is the rotor radius, and $C_\mathrm{T}=4a(1-a)$ is the thrust coefficient with a flow induction factor $a$. Here $x$ is the downstream distance with respect to the turbine.

If several turbines are located upstream of a given turbine of interest, their wake effects accumulate. It was proposed by Kat\'ic {\it et al.}\ \cite{kat86} (also by Lissaman in 1979 \cite{lis79}) that the kinetic energy deficit of the mixed wake is the same as the sum of the energy deficits of upstream wakes that are modeled as if they were each exposed to the unperturbed free-stream velocity $u_{0}$. Thus, Kat\'ic {\it et al.}\ \cite{kat86} proposed to model the wake effects by adding the squared velocity deficits of the individual wakes. The velocity deficit at position ${\bf x}=(x,y,z)$ due to some upstream turbine (turbine $j$) centered at position $(x_\mathrm{j},y_\mathrm{j},z_{\mathrm{h}})$, where $z_{\mathrm{h}}$ is the turbine hub-height, is defined according to
\begin{equation}
\delta u({\bf x};j) = u_{0} - u({\bf x};j) = \frac{2 ~a ~u_{0}}{[1+k_\mathrm{w} (x-x_\mathrm{j}) / R]^2}.
\end{equation}
A non-zero velocity deficit exists only at positions {\bf x} such that there exists an upstream turbine that generates a wake there. Specifically, if the following condition holds:
\begin{equation}
(y-y_\mathrm{j})^2+(z-z_{\mathrm{h}})^2 \leq [R+k_\mathrm{w}(x-x_\mathrm{j})]^2 \mbox{~~~~~~~~~~~~ for ~~~~~~~ $x > x_\mathrm{j}$}.
\label{eq-cond}
\end{equation}
Here $(x-x_\mathrm{j})$ indicates the downstream distance, $(y-y_\mathrm{j})$ the `transverse', and $(z-z_{\mathrm{h}})$ the vertical distance with respect to the turbine hub-height $z_{\mathrm{h}}$. 

The interaction of the wakes with the ground is modeled by incorporating ``ghost'' or ``image'' turbines under the ground surface based on the procedure in Lissaman \cite{lis79}. That is to say, to each turbine $j$ at position $(x_\mathrm{j},y_\mathrm{j},z_{\mathrm{h}})$ we associate an image turbine at $(x_\mathrm{j},y_\mathrm{j},-z_{\mathrm{h}})$. The interaction of the wakes originating from the ``ghost'' turbines with the wakes originating from the actual turbines is assumed to model the reduced rate of wake recovery (and thus larger velocity deficit) due to ground effects. Thus it is assumed that the following two types of upstream wakes interact when modeling the velocity at some turbine location ${\bf x}$:\\
\\
(1) Turbines and underground ``ghost'' turbines directly upstream of point ${\bf x}$ \cite{jen83} (the set of turbines, denoted $J_U$, that are in front of the point ${\bf x}$),\\
(2) Turbines and underground ``ghost'' turbines in adjacent rows whose wakes grow sufficiently to overlap with position ${\bf x}$ (denoted as turbine set $J_S$).\\

\noindent The corresponding superposition of velocity defect kinetic energies leads to to the following model for the velocity at the point ${\bf x}$
\begin{equation}
u({\bf x}) = u_{0} - \sqrt{ \sum_{j \in J_\mathrm{A} } [\delta u({\bf x};j)]^2 },
\label{eq-velsuperposed}
\end{equation}
where 
$J_\mathrm{A}=J_\mathrm{U} \cup J_\mathrm{S} $
 is the union of the two sets of wake effects that are seen at point ${\bf x}$ according to the condition in equation \eqref{eq-cond}. Next, consider points that are located on the rotor disk of a particular wind-turbine $T$. We discretize the disk using a rectangular lattice with an uniform spacing of $6$ meters, which results in about $200$ points per disk, as we consider turbines with a diameter of $100$ meters. The mean velocity at a particular point ${\bf x}_\mathrm{T,k}$ (with $k=1,2,..N_d$) on the turbine disk is given by evaluating equation \eqref{eq-velsuperposed} at the position ${\bf x} = {\bf x}_\mathrm{T,k}$. The ratio of the velocity at that point divided by the incoming unperturbed velocity $u_{0}$ is thus given by
\begin{equation}
\frac{u({\bf x}_\mathrm{T,k})}{u_{0}} = 1 - 2a \sqrt{ \sum_{j \in J_\mathrm{A,k}} [1+k_\mathrm{w} (x_\mathrm{T,k}-x_\mathrm{j}) / R]^{-4} }.
\label{eq-velsuperposed2}
\end{equation}
Note that in this equation the set $J_\mathrm{A,k}$ depends on the specific point ${\bf x}_\mathrm{T,k}$ since different locations on the disk may intersect different wakes from different sets of upstream turbines. The velocity of turbine $T$ with respect to the incoming wind is obtained by computing the average velocity over all points in the turbine disk area using
\begin{equation}
\frac{u_\mathrm{T}}{u_{0}}=\frac{1}{N_\mathrm{d}}\sum_{k=1}^{N_\mathrm{d} } \frac{u({\bf x}_\mathrm{T,k})}{u_{0}}.
\label{eq-velsuperposed3}
\end{equation}
The power $P_\mathrm{T}$ of that turbine normalized with the power of a free-standing turbine $P_\mathrm{1}$ (or the first row of the wind-farm) is given by
\begin{equation}
\frac{P_\mathrm{T}}{P_\mathrm{1}} = \left( \frac{u_\mathrm{T}}{u_{0}} \right)^3. 
\label{eq-ratiowake}
\end{equation}

\begin{figure}
\centering
\subfigure{\includegraphics[width=0.99\textwidth]{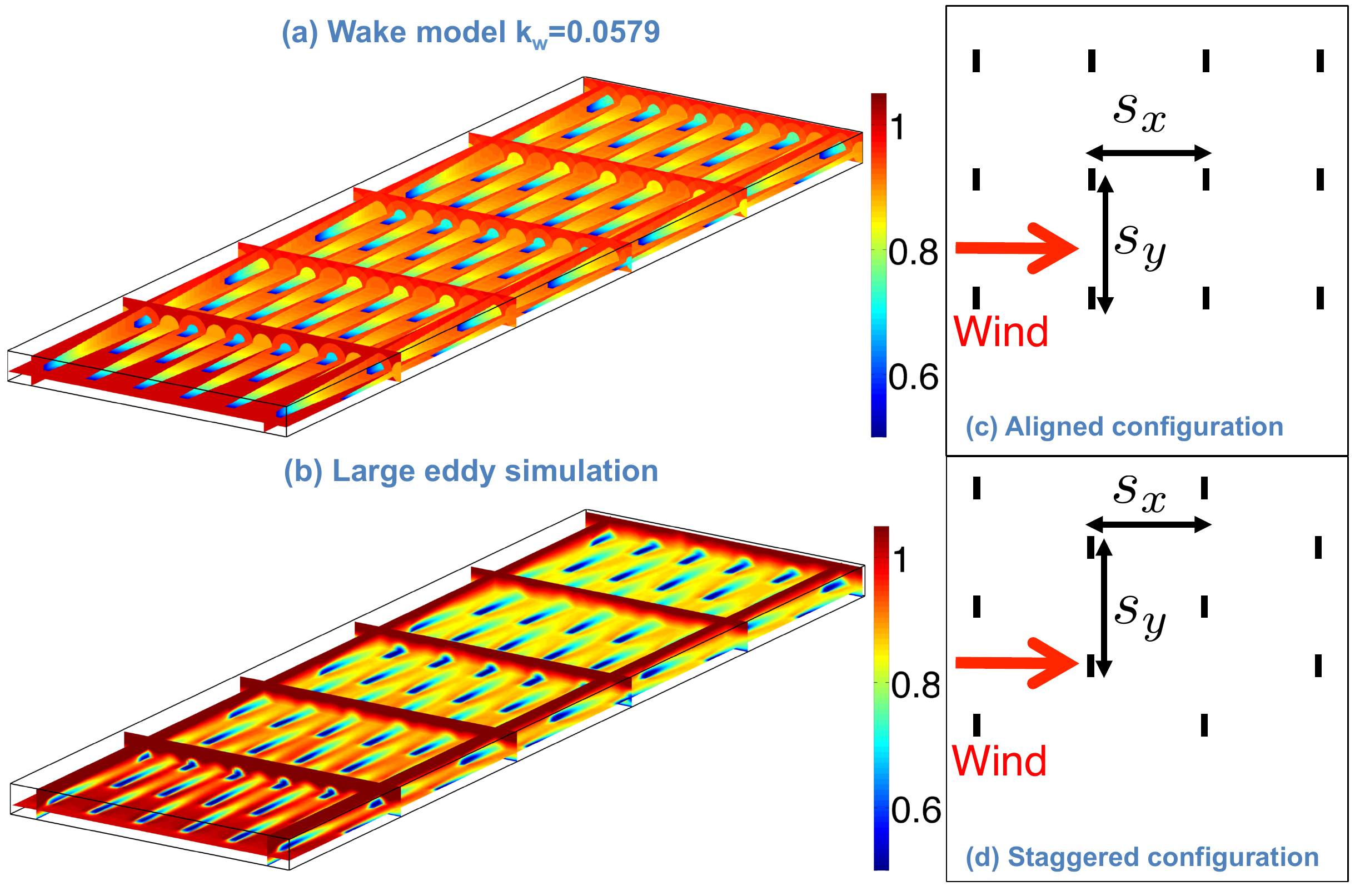}}
\caption{(a,b) Three-dimensional renderings of the mean streamwise velocity in a staggered wind-farm with streamwise spacing $s_\mathrm{x}=7.85$ and a spanwise spacing of $s_\mathrm{y}=5.24$. Panel (a) shows the result from the wake model including axisymmetric linear wake expansion and wake superposition, while (b) shows the results from LES for the same wind-turbine arrangement. See Ref.\ \cite{ste14f} for details about the simulation. Panel (c) and (d) show a sketch of the aligned and staggered configuration, respectively.}
\label{figure2}
\end{figure}

In this model the wind-speed reduction at a particular turbine $T$ is therefore a function of (I) the assumed spatial distribution of the upstream and adjacent turbines, and (II) the wake decay parameter $k_\mathrm{w}$. Frandsen \cite{fra92} proposed a relationship between this parameter and the atmospheric turbulence characteristics. Following a reasoning that was also articulated in Lissaman \cite{lis79}, the growth rate can be assumed to be on the order of the ratio of transverse velocity fluctuations to the mean velocity. Assuming that the former is on the order of the friction velocity, the ratio defining the wake decay parameter becomes
\begin{equation}
k_\mathrm{w}=\frac{\kappa} {\ln(z_{\mathrm{h}}/z_\mathrm{0,lo})},
\end{equation}
where $z_{\mathrm{h}}$ is the height of the turbine, $z_\mathrm{0,lo}$ is the roughness length of the ground surface, and $\kappa=0.4$ is the von K\'arm\'an constant. With this assumed wake coefficient the wake model can be shown to capture the velocity deficits in the beginning of the wind-farm quite well. However, the fully developed regime is not necessarily described well with $k_\mathrm{w}$, see also Refs.\ \cite{bar09b,son14,has09,sch09,bea12,bro12}. As will be shown later, an important ingredient of the coupled model is to adjust the wake expansion coefficient in the fully developed regime of the wind-farm based on parameters obtained from the ``top-down'' model. 

The turbine velocity and power output of the turbines in the fully developed region of the wind-farm can be obtained by applying the wake model to predict the streamwise velocity field $u({\bf x})$ at all points on a three-dimensional mesh. For the calculation presented here we use a resolution of $\Delta x =\Delta y =\Delta z = 6$ meters. To determine the velocity field in the fully developed regime we consider the effect of a very large number of upstream rows. Specifically, we consider 100 upstream rows with up to $4$ columns of turbines on the left and the right side including the corresponding ``ghost'' turbines. These parameters can be shown to lead to fully converged results for the wake model. That is to say, adding more turbines upstream or to the sides does not make any difference in the results. In fact, for most cases only a fraction of the turbines used in this study are necessary to reach convergence. Note that since the model is analytical the full solution only has to be calculated for visualization purposes, see figures \ref{figure2}, and figures \ref{figure14}-\ref{figure17}. To determinate the turbine power outputs the velocity only needs to be calculated at the turbine locations. In appendix 1 we present some practically relevant simplifications that can be used for calculating the velocity at wind-turbine locations more efficiently in the aligned and staggered configurations.

Both the presented LES and model results assume that all turbines operate in region II, where the thrust coefficient $C_\mathrm{T}$ is constant as function of the wind-speed. Note that the experimental data to which we compare the model results in section \ref{Section_HornsRev} are obtained for a wind-speed of $8\pm0.5$m/s, which corresponds to the turbines operating in region II. At the end of section \ref{Section_combined_b} we explain how the CWBL model can be applied to the cases where the $C_\mathrm{T}$ coefficient of the turbines in the wind-farm changes in the entrance region of the wind-farm. In addition, both the LES and model neglect the variation of the power coefficient $C_\mathrm{P}$ with wind-speed. For the comparison with the field experiments presented here this is a reasonable assumption since the data have been obtained for a very narrow range of wind speeds and hence $C_\mathrm{P}$ cancels out when relative turbine power outputs are considered.

Figure \ref{figure2}(a) gives a three-dimensional representation of the predicted mean velocity in a fully developed staggered wind-farm for a dimensionless streamwise spacing (in units of rotor diameter $D$) of $s_\mathrm{x}=7.85$ and a spanwise spacing of $s_\mathrm{y}=5.24$. The geometric average of the spacing is defined as $s=\sqrt{s_\mathrm{x} s_\mathrm{y}}$ and is $s=6.41$ in this case. In figure \ref{figure2}(b) the results from a corresponding LES run, averaged in time, are also shown for a qualitative comparison. The parameters for both the LES and the wake model used here are: $D=100$m and $z_{\mathrm{h}}=100$m. The surface roughness height in the LES was $z_\mathrm{0,lo}=0.1$m and $C_\mathrm{T}=0.75$.

\begin{figure}
\centering
\subfigure{\includegraphics[width=0.999\textwidth]{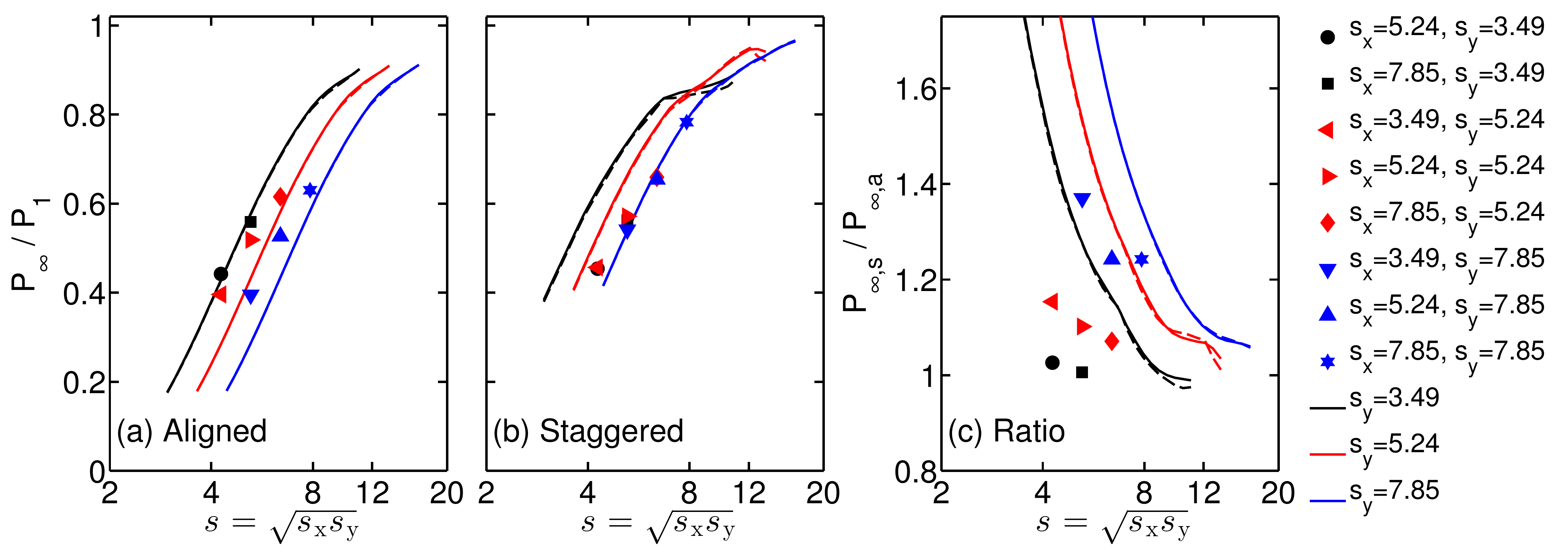}}
\caption{Power output ratio $P_\mathrm{\infty}/P_\mathrm{1}$ in the fully developed regime according to the wake model (equations \eqref{eq-velsuperposed2}-\eqref{eq-ratiowake}) with $k_\mathrm{w}=0.0579$ as function of the geometric mean turbine spacing $s=\sqrt{s_\mathrm{x}s_\mathrm{y}}$. $P_\mathrm{\infty}$ is the power output per turbine in the fully developed regime while $P_\mathrm{1}$ is the reference power output of a single turbine without effects from other turbines. The panels show the model results (lines) for the (a) aligned and (b) staggered configuration compared to LES results (symbols) \cite{ste13,ste14b,ste14f}. Panel (c) shows the ratio between the staggered and aligned results. The dashed lines indicate the sum approximations (equations \eqref{appendix1-1}-\eqref{appendix1-6}) for the wake model given in appendix 1. Later on, Figures \ref{figure5} and \ref{figure9} compare results from the ``top-down'' and CWBL model with LES data. As will be seen, only the proposed CWBL model captures the trend for both the aligned and staggered cases.}
\label{figure3}
\end{figure}

Next, in order to highlight some advantages and drawbacks of the wake model, it is applied (without coupling with the ``top-down'' model) to predict wind-turbine power output for various wind-farm configurations consisting of different streamwise and spanwise turbine spacings. In figure \ref{figure3} the wake model results are compared with the LES results. For the LES the power ratio $P_\mathrm{\infty} / P_\mathrm{1}$ is determined by measuring $\langle \overline{u_{\infty}^3} \rangle / \langle \overline{u_1^3} \rangle$, where $u_\infty$ is the velocity averaged over a turbine disk for turbines at the end of the wind-farm and $u_1$ is the velocity averaged over the disk for turbines in the first row, where the overbar indicates time averaging. We have verified from the LES that the difference in results using this model versus the $\langle \overline{u_\infty / u_1} \rangle^3$ implied in the wake model are negligible. The actual power will be higher using $\langle \overline{u^3} \rangle$ than using $\langle \overline{u} \rangle^3$ due to the fluctuations. However, for the power ratio $P_\mathrm{\infty} / P_\mathrm{1}$ most of these differences cancel out due to the normalization.

Figure \ref{figure3} shows that the model predicts correctly that $P_\mathrm{\infty}/P_\mathrm{1} \to 1$ as $s \to \infty$. In these and all remaining figures the model results are shown for nondimensional streamwise spacings that vary between $s_\mathrm{x}=2.5$ and $s_\mathrm{x}=35$. Figure \ref{figure3} also shows that for aligned wind-farms with the same geometric mean turbine spacing $s$, the power is greater for the cases in which the streamwise distance $s_\mathrm{x}$ is increased while the spanwise distance $s_\mathrm{y}$ is smaller. The LES results (shown as symbols) yield similar trends. Conversely, for the staggered arrangement, the LES results show that all cases tend to collapse onto a single curve, i.e.\ the dependence is mainly on the geometric mean spacing $s$ \cite{ste14f}. The results in figure \ref{figure3}(b) indicate that for the staggered configuration, the wake model does not accurately represent the power output in the fully developed region of the wind-farm. Figure \ref{figure3}(c) compares the relative power output in the fully developed regime for the aligned and staggered configuration and reveals that the differences are largest when the streamwise turbine spacing is small. The importance of this effect is over predicted by the wake model.

\section{``top-down'' model} \label{Section_topdown}
The ``top-down'' wind-farm model traces its origins to Lissaman \cite{lis79}. It was further developed and presented in an updated form by Frandsen \cite{fra92,fra06}. The model is a single-column model of the atmospheric boundary layer based on momentum theory. It postulates the existence of two constant momentum flux layers, one above the turbine hub-height and one below. Each has a characteristic friction velocity and roughness length. Detailed analysis and comparisons with LES \cite{cal10} showed that the assumption inherent in the Frandsen derivation, namely that two logarithmic layers would meet at hub-height needed to be corrected in order to account for the horizontally averaged effects of turbine wakes. The ``top-down'' model by Calaf {\it et al.} \cite{cal10} accounts for such a layer by increasing the eddy-viscosity in this region. This augmented model was shown to predict roughness heights that agree well with results from LES. In this section we first describe this ``top-down'' model in section \ref{Section_topdown1}. Subsequently we discuss in \S \ref{Section_topdown2} the specific role of spanwise spacing in the ``top-down'' model and how the wake model can be used to determine it.

\subsection{Model description} \label{Section_topdown1} 
The objective of the ``top-down'' model is to predict the horizontally and time averaged velocity profile $\langle \overline{u} \rangle(z)$ in the wind-turbine array boundary layer ,where the overbar indicates time averaging. The presentation below follows closely that of Ref.\ \cite{men12} and is included for completeness. The model assumes the presence of two constant stress layers, one above and one below the turbine region \cite{fra92,fra06,cal10,men12,ste14c}. First, as a reference, if there is no wind-farm, then the flow can be assumed to be undisturbed, and we have the traditional logarithmic law:

\begin{align}
\label{Eq_profile_5}
\langle \overline{u_\mathrm{0}} \rangle (z) = \frac{u_{*}}{\kappa} \ln \left( \frac{z}{z_{\mathrm{0,lo} }} \right) & ~~~~~~~~~~ \mbox{for} & z_{\mathrm{0,lo}} & \leq z \leq \delta,
\end{align}
above a surface with roughness length $z_{\mathrm{0,lo}} $ and friction velocity $u_*$. In the cases with a wind-farm, a logarithmic region above the wind-turbine array is characterized by an upper friction velocity $u_{\mathrm{*hi}}$ and the lower logarithmic region by a friction velocity $u_{\mathrm{*lo}}$. Next, one considers the horizontally averaged momentum balance, in which the vertical momentum flux above each turbine in the array (see figure \ref{figure6}) is equal to the stress times the area,
$u_{\mathrm{*hi}}^2 (s_\mathrm{x} s_\mathrm{y} D^2)$. Also, the vertical momentum flux below the turbine is equal to
$u_{\mathrm{*lo}}^2 (s_\mathrm{x} s_\mathrm{y} D^2)$. In the fully developed region of the wind-farm the difference between these two quantities must be the thrust force at the turbine, which is modeled using the thrust coefficient $C_\mathrm{T}$ and the horizontally averaged mean velocity at hub-height $\langle \overline{u}\rangle (z _{\mathrm{h}})$ according to $\frac{1}{2} C_\mathrm{T} [\langle \overline{u}\rangle (z_{\mathrm{h}})]^2 \frac{\pi}{4} D^2$. As a result, we can write 
\begin{equation}
\label{Eq_momentum_balance}
u_{\mathrm{*hi}}^2= u_{\mathrm{*lo}}^2+ \frac{1}{2}c_{\mathrm{ft}} [\langle \overline{u} \rangle (z_{\mathrm{h}})]^2, 
\end{equation}
where $c_{\mathrm{ ft}} = \pi C_\mathrm{T} /(4s_\mathrm{x}s_\mathrm{y})$. 

The modeling of the momentum flux using an appropriate eddy-viscosity allows one to write an equation for the mean velocity $(\kappa ~z~ u_{\mathrm{*lo}} ) d \langle \overline{u} \rangle /dz=u_{\mathrm{*lo}}^2$ inside an assumed constant flux layer below the turbine area that can be integrated from the ground up to yield:
\begin{align}
\label{Eq_profile_1}
\langle \overline{u} \rangle (z) = \frac{u_{\mathrm{*lo}}}{\kappa} \ln \left( \frac{z}{z_{\mathrm{0,lo} }} \right) & ~~~~~~~~~~\mbox{for} & z_{\mathrm{0,lo}} & \leq z \leq z_{\mathrm{h}}-\frac{D}{2},
\end{align}
A similar integration of $(\kappa ~z~ u_{\mathrm{*hi}}) d \langle \overline{u} \rangle /dz=u_{\mathrm{*hi}}^2$ in the layer above the turbine area in which one assumes a roughness length $z_{\mathrm{0,hi}}$ representing the entire wind-farm yields
\begin{align}
\label{Eq_profile_4}
 \langle \overline{u} \rangle (z) = \frac{u_{\mathrm{*hi}}}{\kappa} \ln \left( \frac{z}{z_{\mathrm{0,hi}} } \right) &~~~~~~~~~~ \mbox{for} & z_{\mathrm{h}}+\frac{D}{2} & \leq z \leq \delta,
\end{align}
where $\delta$ is the upper scale, which in the fully developed boundary layer case is on the order of the height of the atmospheric boundary layer (here the ``top-down'' model is only used to model the fully developed region of the wind-farm, although generalizations to the developing case are possible \cite{men12,ste14c}). Inside the wake region $z_{\mathrm{h}}-D/2 \leq z \leq z_{\mathrm{h}}+D/2$ and the horizontally averaged velocity profiles can be obtained by assuming that the eddy viscosity is increased by an additional wake eddy viscosity $\nu_\mathrm{w}$. This gives
\begin{align}
\label{Eq_profile_4b}
(\kappa z u_* + \nu_\mathrm{w}) \frac{d \langle \overline {u} \rangle}{dz} = u_*^2 ~~~~~~ \rightarrow ~~~~~~ (1 + \nu_\mathrm{w}^*) \frac{d \langle \overline {u}\rangle }{d \ln (z / z_{\mathrm{h}})} = \frac{u_*}{\kappa} & ~~~~~~~~~~~~~~\mbox{for} & z_{\mathrm{h}}-{\frac{\mbox{D}}{2} } < z < z_{\mathrm{h}}+\frac{D}{2}.
\end{align}
where $\nu_\mathrm{w}^*= \nu/(\kappa u_*z) \approx \sqrt{ \frac{1}{2} c_{\mathrm{ft}} }~ \langle \overline {u}(z_{\mathrm{h}}) \rangle ~D/ (\kappa u_* z_{\mathrm{h}})$. Since the value of $\nu_\mathrm{w}^*$ depends on the roughness height and the downstream position in the wind-farm, this value should in principle be determined by iteration \cite{ste14c}. In the wake layer the friction velocity is assumed to be $u_{\mathrm{*lo}}$ for $z<z_{\mathrm{h}}$ and $u_{\mathrm{*hi}}$ for $z>z_{\mathrm{h}}$. Vertically integrating this wake layer, and matching the velocities at $z=z_{\mathrm{h}}-D/2$ and $z=z_{\mathrm{h}}+D/2$ gives
\begin{align}
\label{Eq_profile_2}
 \langle \overline{u} \rangle (z) = \frac{ u_{\mathrm{*lo}}}{\kappa} \ln \left[ \left( \frac{z}{z_{\mathrm{h}}} \right)^{\frac{1}{1+\nu_\mathrm{w}^*}} \left( \frac{z_{\mathrm{h}}}{z_{\mathrm{0,lo}} } \right) \left( 1 - \frac{D}{2z_{\mathrm{h}}} \right)^{\beta} \right] 	& ~~~~~~~~~~~~~~\mbox{for}&z_{\mathrm{h}}-\frac{D}{2} &\leq z \leq z_{\mathrm{h}},
\end{align}
and
\begin{align}
\label{Eq_profile_3}
 \langle \overline{u} \rangle (z) = \frac{ u_{\mathrm{*hi}}}{\kappa} \ln \left[ \left( \frac{z}{z_{\mathrm{h}}} \right)^{\frac{1}{1+\nu_\mathrm{w}^*}} \left( \frac{z_{\mathrm{h}}}{z_{\mathrm{0,hi}}} \right) \left( 1 + \frac{D}{2z_{\mathrm{h}}} \right)^{\beta} \right] 	& ~~~~~~~~~~~~~~\mbox{for} &z_{\mathrm{h}} 		 &\leq z \leq z_{\mathrm{h}}+\frac{D}{2}.
\end{align}
In both \eqref{Eq_profile_2} and \eqref{Eq_profile_3} the exponent $\beta$ is defined as $\beta=\nu_\mathrm{w}^*/(1+\nu_\mathrm{w}^*)$. Enforcing continuity between equation \eqref{Eq_profile_2} and \eqref{Eq_profile_3} at $z=z_{\mathrm{h}}$ gives
\begin{equation}
\label{Eq_model_continuity}
\frac{u_{\mathrm{*hi}}}{u_{\mathrm{*lo}}}= \left( \ln\frac{z_{\mathrm{h}}}{z_{\mathrm{0,lo}}} + \beta \ln\left[1-\frac{D}{2z_{\mathrm{h}}}\right] \right) / \left( \ln \frac{z_{\mathrm{h}}}{z_{\mathrm{0,hi}}} + \beta \ln\left[1+\frac{D}{2z_{\mathrm{h}}}\right] \right) . 
\end{equation}
Substituting this relationship into the momentum balance (equation\ \eqref{Eq_momentum_balance}) and replacing the mean velocity at hub-height one can obtain the roughness height $z_{\mathrm{0,hi}}$, as provided later in the paper (equation \eqref{Eq_defz0hi}). Also, matching the velocity at $z=\delta$ between the wind-farm case and the free atmosphere situation (assuming that at this height the velocity assumes a reference value such as that of the geostrophic wind) one has
\begin{equation}
\label{Eq_model_ushi}
u_{\mathrm{*hi}} = u_* \frac{\ln \left(\delta/z_{\mathrm{0,lo}} \right)} {\ln \left( \delta/z_{\mathrm{0,hi}} \right)}.
\end{equation}
Combining this with equation \eqref{Eq_profile_3} allows us to write the velocity from the ``top-down'' model at hub-height as 
\begin{align}
\label{Eq_model_uhub-height}
\langle \overline{u} \rangle(z_{\mathrm{h}}) = \frac{u_*}{\kappa} \frac{\ln \left(\delta/z_{\mathrm{0,lo}} \right)} {\ln \left( \delta/z_{\mathrm{0,hi}} \right)} ~\ln \left[ \left( \frac{z_{\mathrm{h}}}{z_{\mathrm{0,hi}}} \right) \left(1 + \frac{D}{2z_{\mathrm{h}}} \right)^\beta \right].
\end{align}
The ratio of the mean velocity to the reference case without wind-farms is then given by
\begin{equation}
\label{Eq_veloc}
 \frac{\langle \overline{u} \rangle(z_{\mathrm{h}}) }{\langle \overline{u_\mathrm{0}} \rangle(z_{\mathrm{h}})} = \frac{\ln \left(\delta/z_{\mathrm{0,lo}} \right)} {\ln \left( \delta/z_{\mathrm{0,hi}} \right)} ~\ln \left[ \left( \frac{z_{\mathrm{h}}}{z_{\mathrm{0,hi}}} \right) \left(1 + \frac{D}{2z_{\mathrm{h}}} \right)^\beta \right] \left[ \ln \left( \frac{z_\mathrm{h}}{z_{\mathrm{0,lo}} } \right)\right]^{-1} .
\end{equation}
The corresponding power ratio is given by the ratio of cubed mean velocity at hub-height with wind-turbines compared to the reference case without wind-farms:

\begin{equation}
\label{Eq_Power}
\frac{P_\mathrm{\infty}}{P_\mathrm{1}}=\left( \frac{\langle \overline{u} \rangle(z_{\mathrm{h}}) }{\langle \overline{u_\mathrm{0}} \rangle(z_{\mathrm{h}})} \right)^3.
\end{equation}
\begin{figure}
\centering
\subfigure{\includegraphics[width=0.49\textwidth]{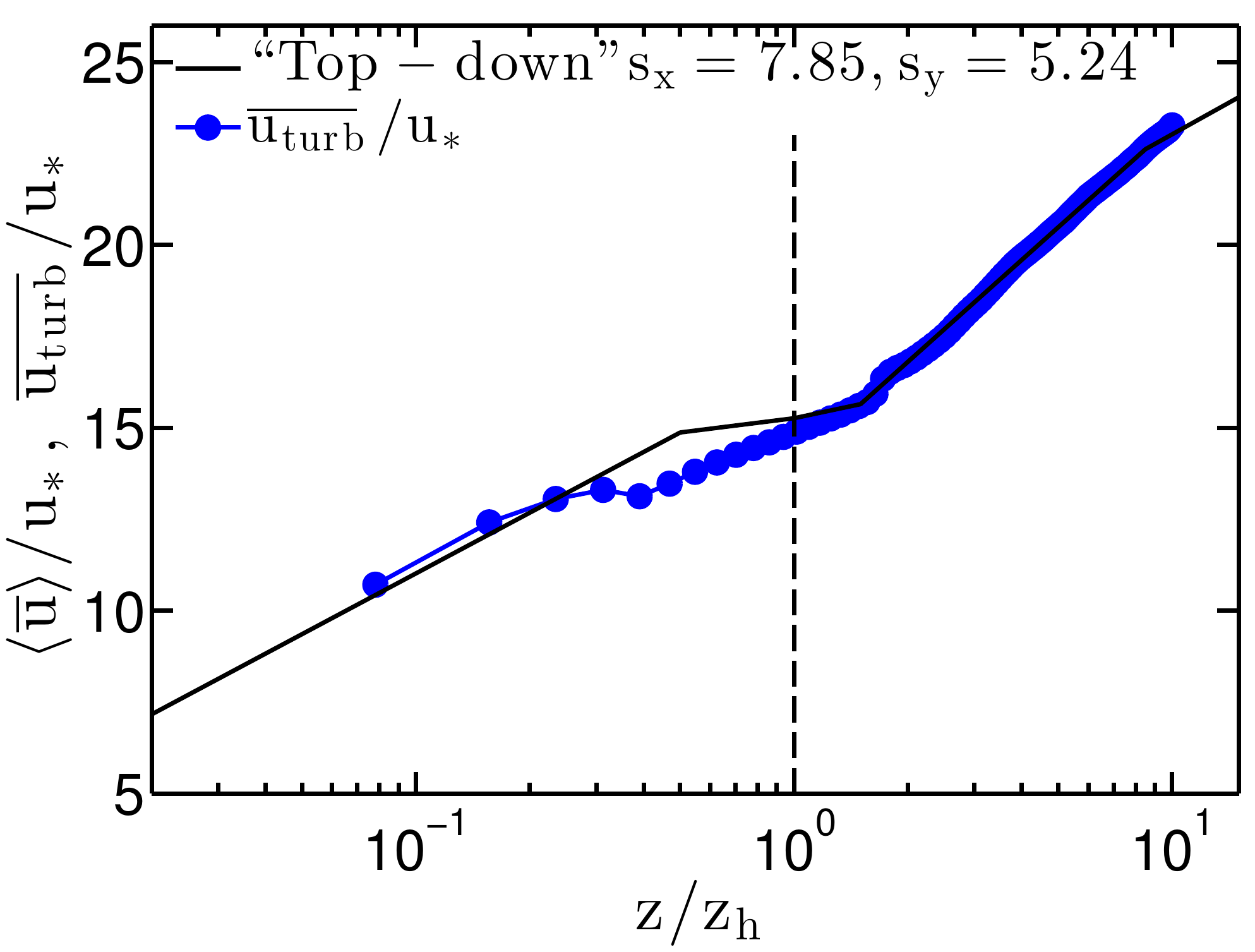}}
\caption{Comparison of the streamwise velocity profile obtained using the ``top-down'' model (solid black line) with the streamwise turbine velocity profile, i.e.\ the velocity in front of the turbines, obtained from an infinitely large staggered wind-farm LES with $s_\mathrm{x}=7.85$ and $s_\mathrm{y}=5.24$. The dashed vertical line indicates the turbine hub-height.}
\label{figure4}
\end{figure}

Figure \ref{figure4} shows a comparison of the streamwise velocity profile obtained from the ``top-down'' model, i.e.\ equations \eqref{Eq_profile_1} - \eqref{Eq_profile_3}, with the streamwise ``turbine'' velocity measured in an infinitely long staggered wind-farm simulation with $s_\mathrm{x}=7.85$ and $s_\mathrm{y}=5.24$ \cite{ste14e}. The figure shows that the ``top-down'' model correctly captures the turbine velocity at hub-height, see details in appendix 2, but does not very accurately capture the velocity near the ground.

\begin{figure}
\centering
\subfigure{\includegraphics[width=0.999\textwidth]{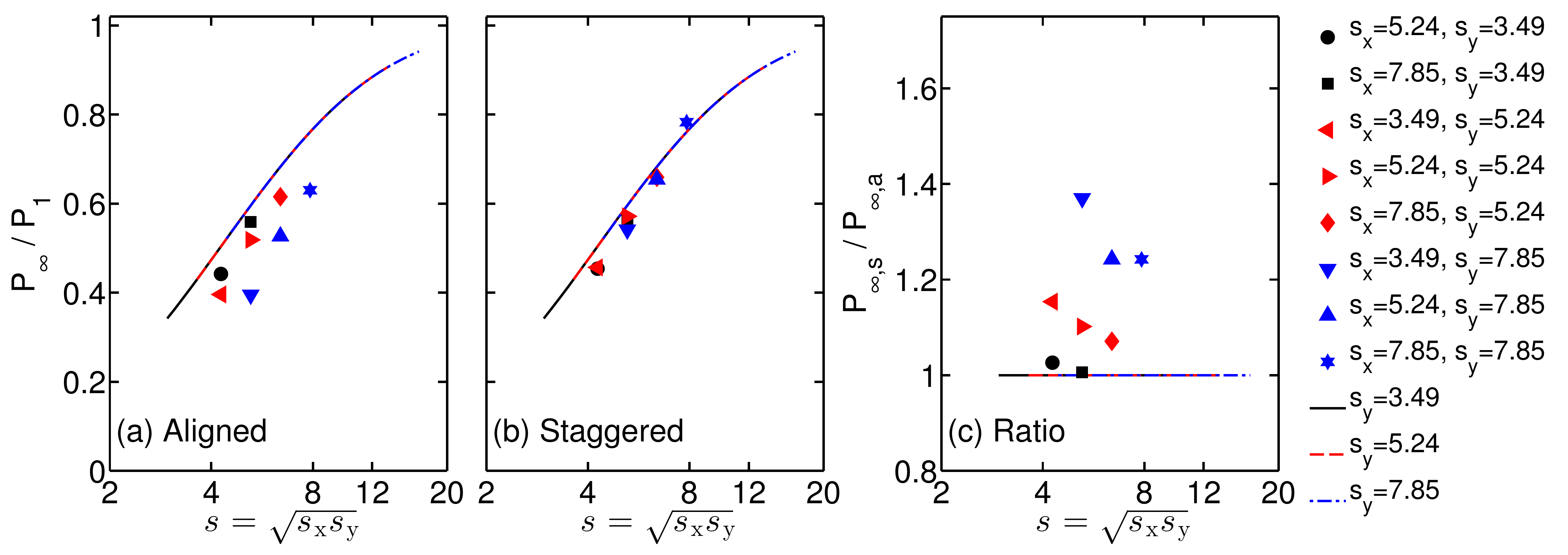}}
\caption{Comparison of the ``top-down'' model (lines) and LES results (symbols) \cite{ste13,ste14b,ste14f} for the relative power output in the fully developed regime ($P_\mathrm{\infty}/P_\mathrm{1}$) in (a) aligned and (b) staggered wind-farms as function of the geometric mean turbine spacing $s=\sqrt{s_\mathrm{x}s_\mathrm{y}}$. Panel (c) gives the ratio between the aligned and staggered case $P_\mathrm{\infty,s}/P_\mathrm{\infty,a}$. Figure \ref{figure3} and figure \ref{figure9} show the comparison of the wake model and the CWBL model to the LES data. Note that only the CWBL model captures the trend for both the aligned and staggered configurations.}
\label{figure5}
\end{figure}

Figure \ref{figure5} compares the ``top-down'' model predictions with results from LES. As expected, the results only depend upon the geometric mean of the turbine spacing ($s$) and no distinction can be made between the aligned and staggered cases. Remarkably, the predictions for the staggered cases appear in very good agreement with the results of Refs.\ \cite{mey12,ste14c}. However, for the aligned cases significant differences can be seen, especially in those cases where the spanwise spacing is large. These large spacings lead to the power degradation being underestimated by the ``top-down'' model. In examining the outputs from the LES, we observe that in the cases in which the spanwise spacing between turbines is large, there is little sideways interactions among the turbines even for the fully developed case. There remains significant spanwise inhomogeneity even in the fully developed case. At large spanwise spacings, the ``top-down'' model is less accurate but its predictions can be improved by including knowledge about the wake expansion, as discussed in the next subsection.

An additional comment about the precise position of matching between the wake and upper log layer is pertinent here. In Calaf {\it et al.} \cite{cal10}, and in Eqs. \eqref{Eq_profile_4} and \eqref{Eq_profile_4b} above, the wake layer is taken to extend up to a height of $z_{\mathrm{h}}+D/2$ where it meets the upper log layer. Conversely, in Stevens \cite{ste14c}, the limit between the two layers was assumed to be at $z_{\mathrm{h}}+D/4$. Our simulations have shown that the latter provides a better fit for the cases of more loaded wind-farms, i.e.\ for smaller turbine spacings $s$ (and/or larger $C_\mathrm{T}$). Conversely, a matching at $z_{\mathrm{h}}+D/2$ provides better predictions for wind-farms with wider spacings. Therefore, better overall predictions could be achieved by specifying that the matching occurs at a height that changes as function of $c_{\rm ft}$, i.e.\ at $z_{\mathrm{h}}+D/q(c_\mathrm{ft})$ where $q(c_{\rm ft})\to 2$ for low $c_\mathrm{ft}$ and $q(c_{\rm ft})\to 4$ at high $c_{\rm ft}$. For the sake of simplicity in this paper we shall proceed with the original formulation of Calaf {\it et al.} \cite{cal10} with the matching at $z_{\mathrm{h}}+D/2$. However, we note that future improvements of the model with more finely tuned parameter dependencies are possible.

\subsection{Effective spanwise spacing in the ``top-down'' model} \label{Section_topdown2}

\begin{figure}
\centering
\subfigure{\includegraphics[width=0.85\textwidth]{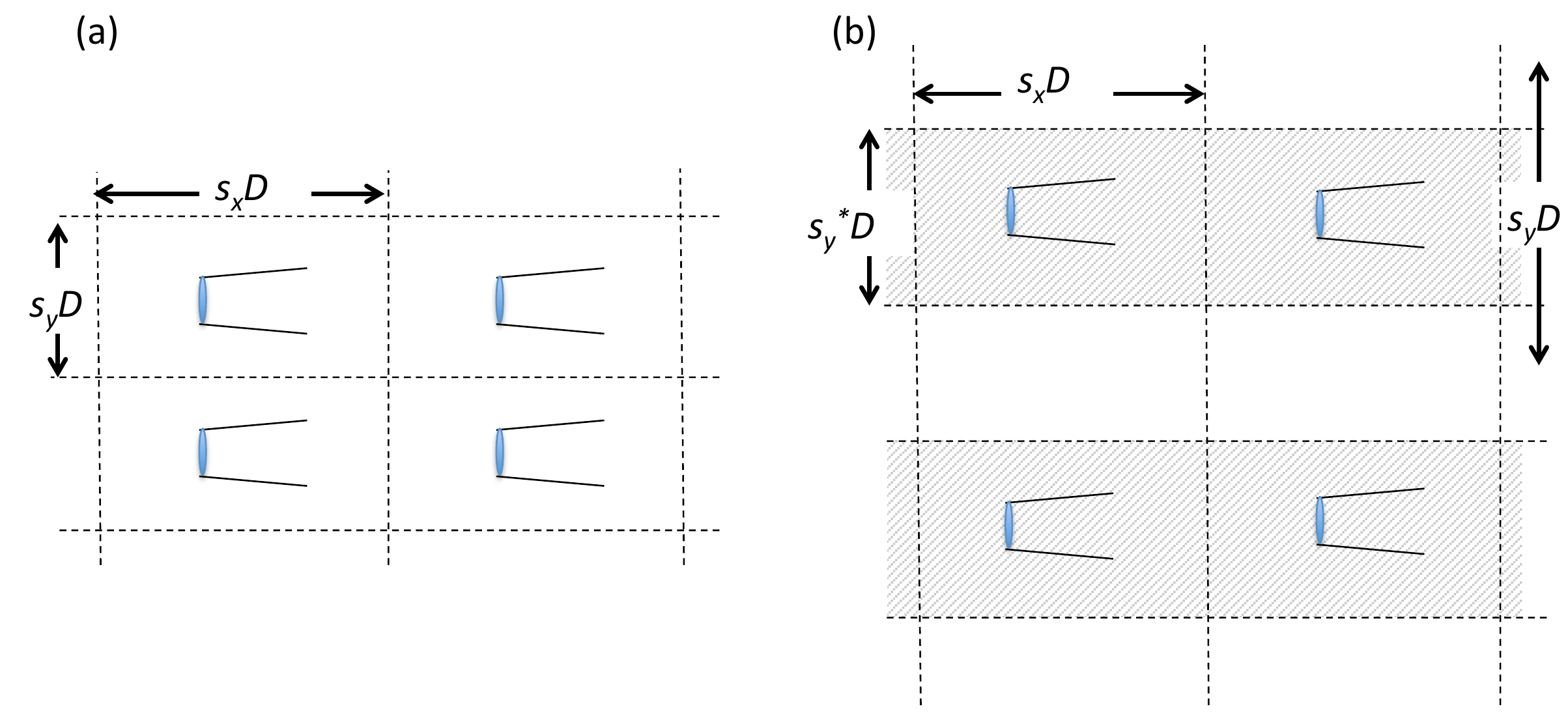}}
\caption{Wind-farm parameters for the ``top-down'' model for the fully developed case and the control volumes used in the momentum analysis. (a) For small spanwise spacings the used control volume coincides with the actual spacing $s_\mathrm{ye}=s_\mathrm{y}$. (b) For widely spaced cases, the control volume (dashed region) uses a smaller spanwise length $s_\mathrm{ye}=s_\mathrm{y}^*<s_\mathrm{y}$ which can be determined using the wake model.}
\label{figure6}
\end{figure}

The LES results indicate that, depending on the spanwise spacing, the velocity deficit due to the turbine wakes can be confined into narrow ``channels''. This confinement is most likely to occur in an aligned configuration, where high velocity wind channels are formed in between the turbine rows. The ``top-down'' model considers a momentum balance averaged over the entire horizontal plane. It thus relates the horizontally averaged velocity with the friction velocity, which depends upon the stresses that are directly affected by the wind-farm near the turbines. However, when the spanwise spacing between turbines becomes larger than some threshold spacing, which we will denote by $s_\mathrm{y}^*$, this assumed association between the mean velocity and the mean momentum fluxes is no longer valid. The limiting case of small $s_\mathrm{x}$ and $s_\mathrm{y} \to \infty$ in the ``top-down'' model that only depends upon $s$ is obviously unrealistic since even for a single line of turbines aligned in the wind direction significant power degradation is to be expected. Hence, we propose to apply the momentum analysis of the ``top-down'' model to a more limited area which is directly affected by the turbine wakes (this region is the shaded area in figure \ref{figure6}). For each wind-turbine the area has length $s_\mathrm{x}D$ as before, but the spanwise length becomes $s_\mathrm{ye}D$ where $s_\mathrm{ye}=\min(s_\mathrm{y},s_\mathrm{y}^*)$ is the ``effective spanwise distance'' between turbines. Then in general, we consider the vertical momentum flux above and below the turbine to be $u_{\mathrm{*hi}}^2 (s_\mathrm{x} s_\mathrm{ye} D^2)$ and $u_{\mathrm{*lo}}^2 (s_\mathrm{x} s_\mathrm{ye} D^2)$ respectively. The thrust force at the turbine has the same expression and thus the momentum balance in this effective region is governed by equation \eqref{Eq_momentum_balance}, with $c_{\rm ft} = \pi C_\mathrm{T} /(4s_\mathrm{x}s_\mathrm{ye})$. 
 
In order to determine $s_\mathrm{y}^*$, information about the strength of spanwise interactions among the turbines is required. Such information is not available within the context of the horizontally averaged ``top-down'' model but it is available from the wake model and is a crucial ingredient in the coupled approach.

\section{The Coupled Wake Boundary Layer (CWBL) model} \label{Section_Combined}

In the previous section we have seen the requirement to determine the effective spanwise spacing $s_\mathrm{ye}$ needed for the ``top-down'' model. In this section we explain the two-way coupling between the ``top-down'' and the wake models. We begin by discussing the fully developed regime in section \ref{Section_combined_a} and extend the approach to the entrance region in section \ref{Section_combined_b}.

\subsection{The fully developed regime} \label{Section_combined_a}

\begin{figure}
\centering
\includegraphics[width=0.99\textwidth]{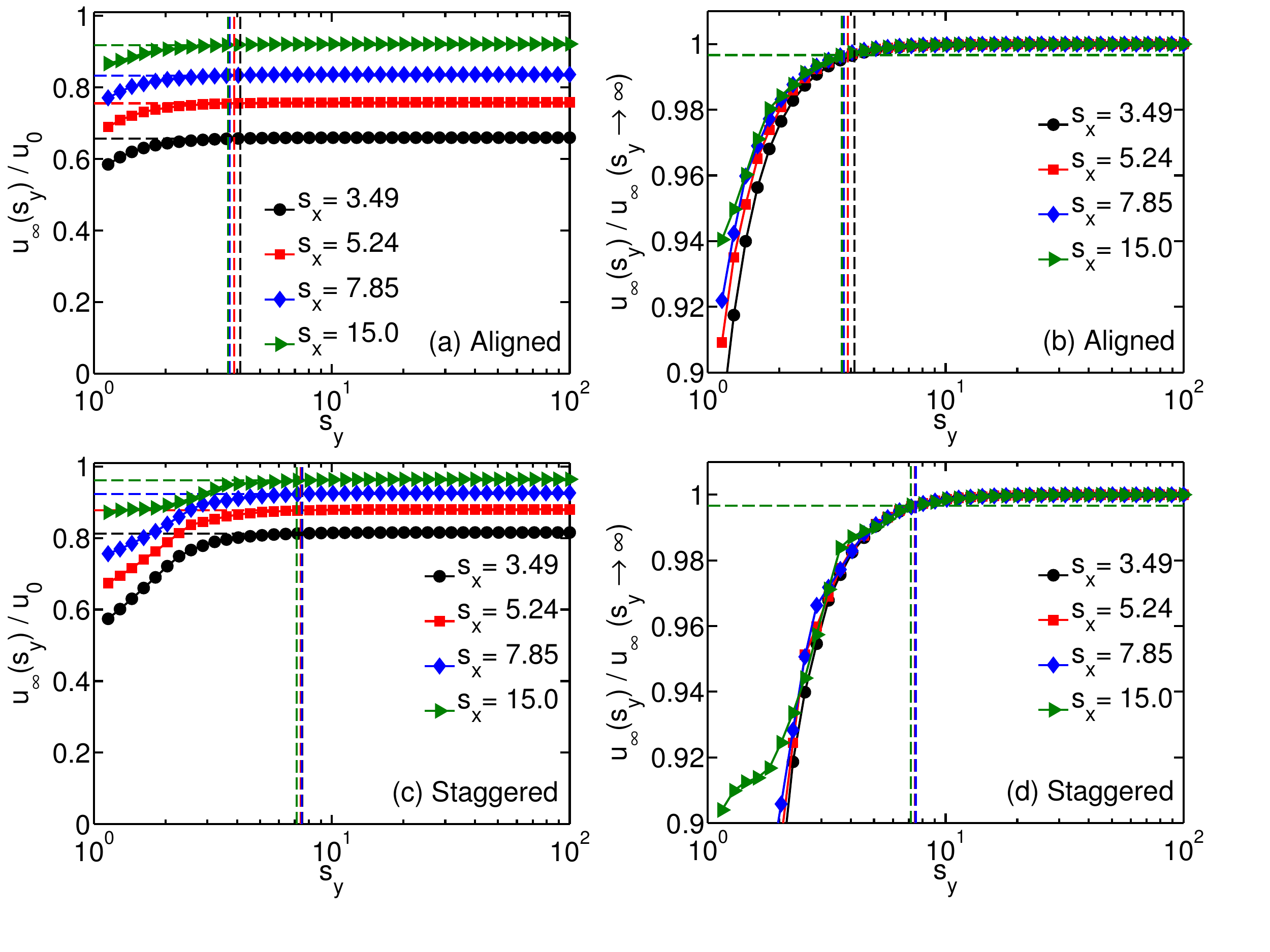}
\caption{Determination of $s_\mathrm{y}^*$ for the aligned (panels (a) and (c)) and staggered (panels (b) and (d)) configuration using $k_\mathrm{w}=0.0579$. Panels (a) and (b) illustrate how the value can be obtained from the turbine velocity in the fully developed regime $u_{\infty}$ as function of $s_\mathrm{y}$ for different streamwise spacings $s_\mathrm{x}$. Panels (c) and (d) show $u_{\infty}(s_\mathrm{y})/u_{\infty}(s_\mathrm{y} \rightarrow \infty)$. In each panel $s_\mathrm{y}^*$ is the spanwise spacing for which $u_{\infty}(s_\mathrm{y})=0.99^{1/3} u_{\infty}(s_\mathrm{y} \rightarrow \infty)$ (dashed horizontal lines) and is indicated by the dashed vertical lines.}
\label{figure7}
\end{figure}

\begin{figure}
\centering
\includegraphics[width=0.99\textwidth]{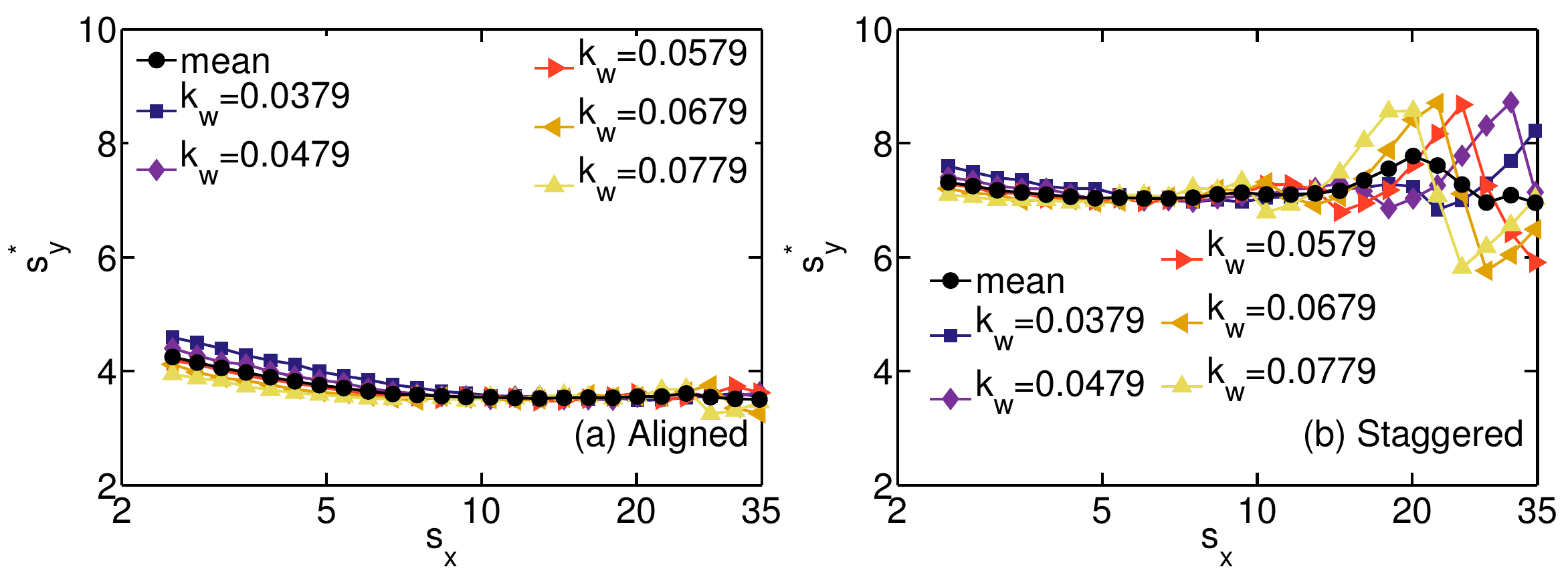}
\caption{$s_\mathrm{y}^*$ for an (a) aligned and a (b) staggered wind-farm as function of the streamwise turbine distance for different wake expansion coefficients $k_\mathrm{w}$.}
\label{figure8}
\end{figure}

A sketch of the coupling between the wake model and the ``top-down'' model is given in figure \ref{figure1}. The procedure requires an initial guess for the wake expansion coefficient $k_\mathrm{w}$ which is used to determine the effective spanwise spacing $s_\mathrm{ye}$ according to the following procesure: $s_\mathrm{y}^*$ is determined from the wake model by finding the spanwise distance for which the spanwise wake effects are negligible, i.e.\ for which the velocity at wind-turbines differs only by $\tfrac{1}{3} \%$ (a fraction of 0.0033) of the velocity obtained for a single line of turbines (the $\tfrac{1}{3}$ \% maps into a $\sim$ 1\% difference in predicted power). To explain the procedure, consider the predictions of the wake model applied for the ``infinite'' (very large) wind-farm for a given $s_\mathrm{x}$ and $s_\mathrm{y}$ shown in figure \ref{figure7}. Note that convergence is obtained due to the fact that wake-wake interactions are modeled by adding the squared velocity deficits, which implies that wakes from turbines far away have a negligible effect on the velocity deficit at a certain point. It is apparent that the turbine velocity increases with $s_\mathrm{x}$ as well as with $s_\mathrm{y}$, but the latter effect saturates after a particular value of $s_\mathrm{y}^*$. For spanwise spacings above this value, the turbine velocities are no longer dependent upon the spanwise spacing. In all of the results presented in this section the $\tfrac{1}{3}$\% ($0.99^{1/3}\sim 0.9967$) threshold is indicated by the dashed lines in each case.

Figure \ref{figure8} shows how $s_\mathrm{y}^*$ depends on the streamwise distance ($s_\mathrm{x}$) and the wake decay coefficient ($k_\mathrm{w}$). Figure \ref{figure8}(a) shows that $s_\mathrm{y}^* \approx 3.5$ for the aligned case and $s_\mathrm{y}^* \approx 7.5$ for the staggered case. We find that $s_\mathrm{y}^*$ depends weakly on the streamwise distance and the wake decay coefficient. Note that especially for the staggered configuration, the values of $s_\mathrm{y}^*$ do not collapse to a single curve for large turbine spacings. The reason is that for these very large turbine spacings the wakes are very weak and $s_\mathrm{y}^*$ defined based on the threshold can vary significantly, especially when plotted as function of the logarithm of the streamwise spacing. In the limit of large $s_\mathrm{x}$, the predictions are almost independent of spacing, hence these features do not have noticeable impact in practice. 
 
The effective spanwise spacing $s_\mathrm{ye}= \min{\left(s_\mathrm{y},s_\mathrm{y}^*\right)}$ from the ``top-down'' model is used to predict the mean horizontal velocity at hub-height, normalized by the reference inflow velocity ${\langle \overline{u} \rangle }/{\langle \overline{u_\mathrm{0}} \rangle }$ at $z_{\mathrm{h}}$ according to equation \eqref{Eq_veloc}. Using the same initial guess for the wake expansion coefficient, the wake model is used to predict the velocity ratio using equation \eqref{eq-velsuperposed} applied to turbines in the fully developed regime of the wind-farm. Since the assumed wake expansion coefficient $k_\mathrm{w}$ may not appropriately reflect the asymptotic effects of turbulence in the boundary layer, there is no guarantee that the two predictions will be the same, i.e.\ typically we find that ${\langle \overline{u} \rangle }/{\langle \overline{u_\mathrm{0}} \rangle }(z_{\mathrm{h}}) \neq u_{T,\infty}/u_\mathrm{0}$ for $k_{w}=\kappa/\ln(z_{\mathrm{h}}/z_\mathrm{0,lo})$ using the actual spanwise spacing $s_\mathrm{y}$ in the ``top-down'' model. Note that $u_{T,\infty}/u_\mathrm{0}$ is the turbine velocity at the end of a very large wind-farm in the wake model. Therefore the wake expansion $k_\mathrm{w}$ and the effective spanwise spacing $s_\mathrm{ye}$ are iterated until convergence is reached, see figure \ref{figure1}. 

The details of the CWBL model can be summarized as follows: \\

Begin by assuming a value of the expansion parameter, e.g.\ assume that $k_\mathrm{w,\infty}=k_\mathrm{w,0}$, where $k_\mathrm{w,\infty}$ is the wake expansion coefficient in the wake model in the fully developed regime of the wind-farm and $k_\mathrm{w,0}$ the wake coefficient in the entrance region of the wind-farm.
\begin{enumerate}
\item For the current value of $k_\mathrm{w,\infty}$ determine $s_\mathrm{y}^*$ from the wake model, by finding the value of $s_\mathrm{y}$ that solves 
\begin{equation}
\frac{u_\mathrm{T}}{u_\mathrm{0}}(s_\mathrm{y},s_\mathrm{x},k_\mathrm{w,\infty},\mathrm{layout},,..)= \frac{u_\mathrm{T}}{u_\mathrm{0}}(s_\mathrm{y} \to \infty,s_\mathrm{x},k_\mathrm{w,\infty},\mathrm{layout},..) - \epsilon, 
\end{equation}
where 
\begin{equation}
\frac{u_\mathrm{T}}{u_\mathrm{0}}(s_\mathrm{y},s_\mathrm{x},k_\mathrm{w,\infty},\mathrm{layout},..)= \frac{1}{N_d}\sum_{k=1}^{N_d} \left[1 - 2a \left( \sum_{j \in J_{A,k} } \left[1+k_\mathrm{w,\infty} \frac{x_\mathrm{T,k}-x_\mathrm{j}}{R} \right]^{-4} \right)^{1/2} \right]
\end{equation}
when applying the wake model to a very large wind-farm in the fully developed regime. In practice, the limit $s_\mathrm{y}\to \infty$ is replaced by $s_\mathrm{y}=200$ and the threshold $\epsilon$ is chosen as $\epsilon = 1- 0.99^{1/3} \approx 0.0033$.\\

\item Use the result above to compute $s_\mathrm{ye}=\min(s_\mathrm{y},s_\mathrm{y}^*)$. \\

\item Calculate ${\langle \overline{u} \rangle }/{\langle \overline{u_\mathrm{0}} \rangle}$ at $z=z_{\mathrm{h}}$ with the ``top-down'' model and find the wake expansion coefficient $k_\mathrm{w,\infty}$ that makes it consistent with the wake model. Equating equations \eqref{Eq_veloc} and \eqref{eq-velsuperposed3}, and replacing the expression for $z_\mathrm{0,hi}$ leads to a single equation for $k_\mathrm{w,\infty}$:
\begin{equation} \label{Eq-combinedvel}
\frac{u_\mathrm{T}}{u_\mathrm{0}}(s_\mathrm{ye},s_\mathrm{x},k_\mathrm{w,\infty},\mathrm{layout},..) = 
 \frac{\ln \left(\delta/z_{\mathrm{0,lo}} \right)} {\ln \left( \delta/z_{\mathrm{0,hi}} \right)} ~\ln \left[ \left( \frac{z_{\mathrm{h}}}{z_{\mathrm{0,hi}}} \right) \left(1 + \frac{D}{2z_{\mathrm{h}}} \right)^{\beta} \right] \left[ \ln \left( \frac{z_{\mathrm{h}}}{z_{\mathrm{0,lo}} } \right)\right]^{-1},
\end{equation}
where
\begin{equation} \label{Eq_defz0hi}
z_{\mathrm{0,hi}}= z_{\mathrm{h}} \left(1+\frac{D}{2 z_{\mathrm{h}}}\right)^{\beta} \\
\exp \left(- \left[ \frac{\pi C_\mathrm{T}}{8s_\mathrm{x} s_\mathrm{ye}\kappa^2} + \left( \ln \left[ \frac{z_{\mathrm{h}}}{z_{\mathrm{0,lo}}} \left( 1 -\frac{D}{2z_{\mathrm{h}}}\right)^{\beta}	\right] \right)^{-2}		\right]^{-1/2} 				\right)
\end{equation}
with $\beta=\nu_\mathrm{w}^*/(1+\nu_\mathrm{w}^*)$, and $\nu_\mathrm{w}^*\approx 28 \sqrt{ \frac{\pi C_\mathrm{T}}{8s_\mathrm{x}s_\mathrm{ye}} }$. This estimate for $\nu_\mathrm{w}^*$ was obtained by Calaf {\it et al.}\ \cite{cal10} for $z_{\mathrm{h}}/z_\mathrm{0,lo}=1000$. As indicated before the actual value for $\nu_\mathrm{w}^*$ should be obtained by iteration. However, for simplicity, we use the above approximation as we find that using $\nu_\mathrm{w}^*\approx 28 \sqrt{ \frac{\pi C_\mathrm{T}}{8s_\mathrm{x}s_\mathrm{ye}} }$ seems to give almost the same answer for the ``top-down'' model as is obtained through the iterative procedure. Note that with this approximation the right hand side of equation \eqref{Eq-combinedvel} can be easily evaluated using $\kappa=0.4$ and the appropriate $C_\mathrm{T}$, $z_{\mathrm{h}}$, $s_\mathrm{x}$, $s_\mathrm{ye}$, $z_{\mathrm{0,lo}}$, $D$, and $\delta$ (for the results shown herein the internal boundary layer height $\delta$ is set to the measured value in the LES, i.e.\ $850$m \cite{ste14c}) parameters based on the particular wind-farm. The left hand side, i.e.\ the wake model part of the model, takes the relative turbine positions into account.

\end{enumerate}
We iterate steps 1 to 3 until equation \eqref{Eq-combinedvel} is satisfied to within some prescribed accuracy. For the results shown here we use a tolerance of $0.05\%$.

\subsection{The entrance region of the wind-farm} \label{Section_combined_b}

The entrance region of the wind-farms can be considered by using the wake portion of the coupled model and assuming that the wake expansion coefficient at the entrance of the wind-farm is equal to the free stream value $k_\mathrm{w,0}=\kappa/\log(z_{\mathrm{h}}/z_\mathrm{0,lo})$ (in our case we use $\kappa = 0.4$, $z_{\mathrm{h}}/z_\mathrm{0,lo} = 1000$ for $z_{\mathrm{h}}=100$m, and $z_\mathrm{0,lo} = 0.1$m, i.e.\ $k_\mathrm{w,0}=0.0579$). This approach is chosen as the free stream wake expansion coefficient seems to describe the entrance region of the wind-farm reasonably well. We assume that the wake expansion coefficient merges continuously towards the value of $k_\mathrm{w,\infty}$ found using the analysis presented in \S \ref{Section_combined_a} for the fully developed region of the wind-farm. The following empirical interpolation function is used to determine the expansion coefficient for the turbines in the wind-farm:
\begin{equation}
\label{equation_finitewind-farm}
k_{\mathrm{w,T}}= k_\mathrm{w,\infty}+(k_\mathrm{w,0}-k_\mathrm{w,\infty}) \exp(-\zeta m),
\end{equation}
where $m$ is the number of turbine wakes that overlap with the turbine of interest and $\zeta$ is an empirical parameter determining the rate at which the asymptotic behavior is reached. Based on an analysis of our results a good choice is $\zeta = 1$. Note that this approach means that the wake model part of the model dominates in the entrance region of the wind-farm, while the wake development further downstream is determined by the coupling between the wake and ``top-down'' models.

Note that the CWBL model can also be applied for cases in which $C_\mathrm{T}$ varies as function of mean velocity, which can sometimes occur in the entrance region of a wind-farm. It is important to realize that the coupling between the wake and ``top-down'' models is performed in the fully developed regime of the wind-farm where $C_\mathrm{T}$ can be assumed constant since the turbines all have the same mean velocity. Therefore no specific changes to the CWBL model are necessary to consider the effect of turbines operating at different $C_\mathrm{T}$ values in the entrance region of the wind-farm. The appropriate turbine specific $C_\mathrm{T}$ can be selected in the wake model part of the CWBL model as would be common practice in wake model calculations.

\section{Results} \label{Section_Results}
In this section we compare the predictions of power degradation using the CWBL approach with LES results from Refs.\ \cite{ste13,ste14b,ste14f}. We first focus on the comparisions in the fully developed regime (section \ref{Section_Results_fully}) and in section \ref{Section_Results_finite} we perform a comparison of the model and LES at the entrance of the wind-farm. A more detailed comparison of the downstream development of the entire mean velocity field from both CWBL and LES for several cases is given in section \ref{Section_Details}.

\subsection{The fully developed regime} \label{Section_Results_fully}

\begin{figure}
\centering
\subfigure{\includegraphics[width=0.99\textwidth]{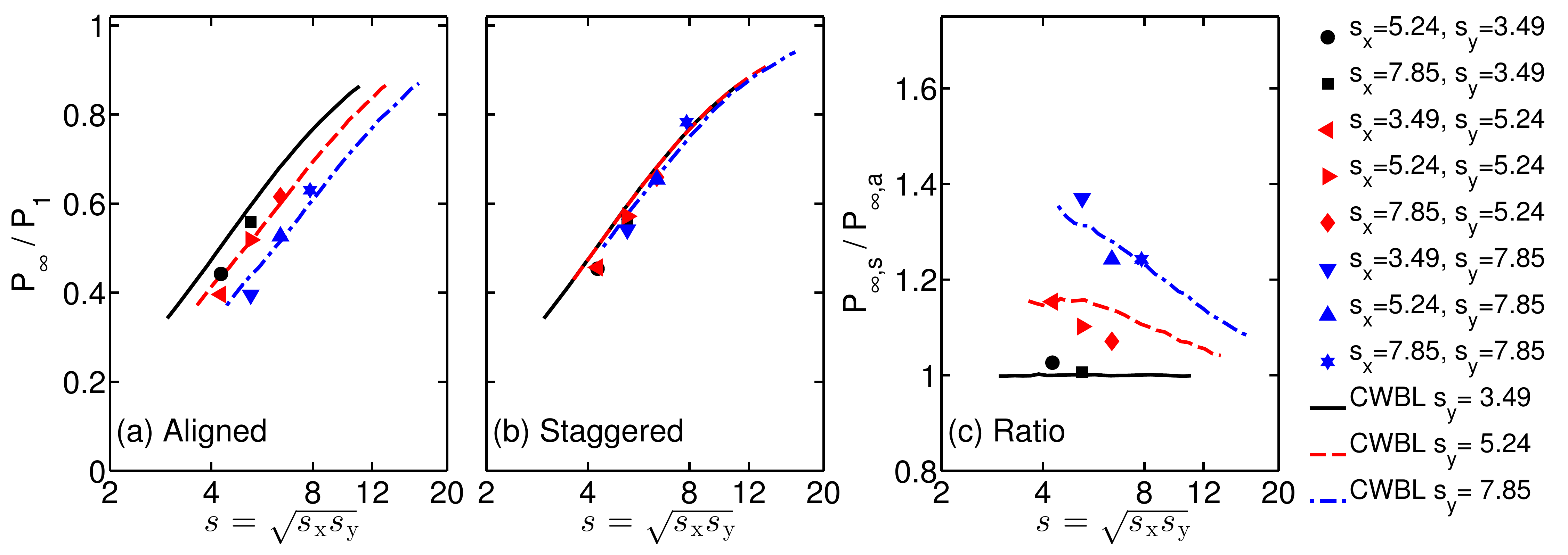}}
\caption{Comparison of the CWBL model (lines) and LES results (symbols) \cite{ste13,ste14b,ste14f} for the relative power output in the fully developed regime ($P_\mathrm{\infty}/P_\mathrm{1}$) in (a) aligned and (b) staggered wind-farms as function of the geometric mean turbine spacing $s=\sqrt{s_\mathrm{x}s_\mathrm{y}}$. Panel (c) gives the ratio between the staggered and aligned case $P_\mathrm{\infty,s}/P_\mathrm{\infty,a}$. Figure \ref{figure3} and figure \ref{figure5} show the comparison of the wake model and the ``top-down'' to the LES data. Note that only the CWBL model captures the trend for both the aligned and staggered configuration.}
\label{figure9}
\end{figure}

\begin{figure}
\centering
\includegraphics[width=0.99\textwidth]{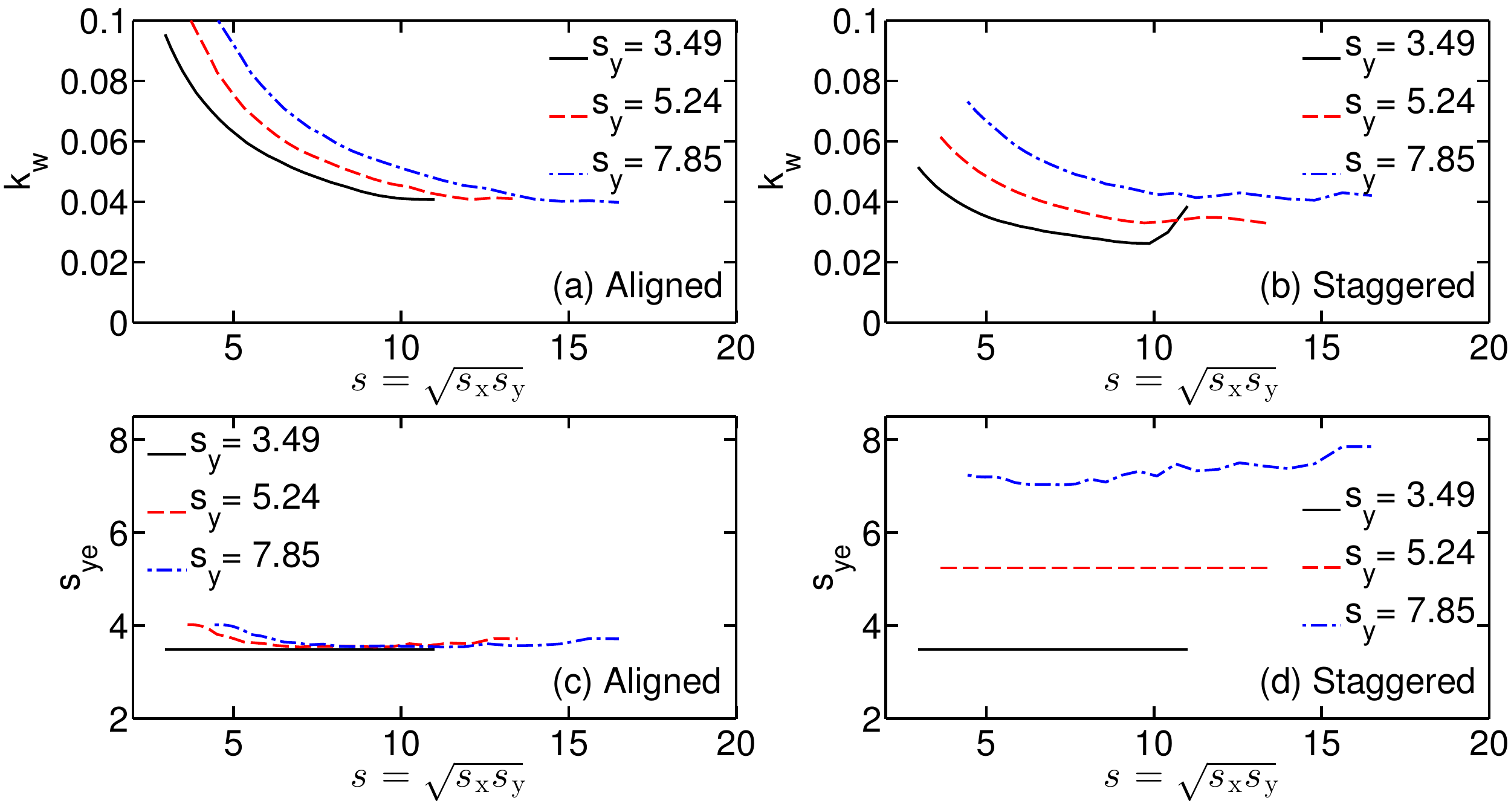}
\caption{Panels (a) and (b) give the wake expansion coefficient $k_\mathrm{w}$ computed from the CBWL model and panel (c) and (d) the corresponding effective spanwise spacing $s_\mathrm{ye}$ as function of the geometric mean turbine spacing $s=\sqrt{s_\mathrm{x}s_\mathrm{y}}$ for the aligned and staggered cases.}
\label{figure10}
\end{figure}

Figures \ref{figure9} compares the power output in the fully developed regime of the wind-farm obtained from LES with the CWBL model results. The figure reveals that the model accurately captures the main trends observed in the LES data. A comparison with figures \ref{figure3} and \ref{figure5} reveals that the CWBL model reproduces the LES data better than the individual, uncoupled models. 

For the wind-farm configurations considered with spanwise spacings up to $\sim8D$, in both the LES and the CWBL model the power output in the fully developed regime depends mainly on the geometric mean turbine spacing when the configuration is staggered, while for the aligned case the ratio between the spanwise and streamwise spacing is also very important \cite{ste14f}. Figure \ref{figure9}c shows that the ratio of the power output of the staggered and the aligned configuration depends on the spanwise spacing. For small spanwise spacings the power output in the fully developed regime is nearly the same in both configurations. A significantly higher power output in the fully developed regime is obtained when the spanwise spacing in the wind-farm is larger than $4D$. 

Figure \ref{figure10}a and \ref{figure10}b show the wake expansion coefficient obtained after the iterations of the CWBL model for the fully developed regime of the wind-farm. The results show that the wake expansion coefficient is larger for the aligned than for the staggered configuration. This means that the wakes are recovering faster when the turbines are aligned compared to when they are staggered. This trend captures the faster wake recovery that has been observed for an aligned wind-farm configuration compared to a staggered one \cite{ste14f}. This faster recovery means that aligned wind-farms with short streamwise turbine spacings perform better than one would expect \cite{ste14f}. 

Note that for large $s_\mathrm{x}$ the $k_\mathrm{w,\infty}$ obtained for the fully developed regime is different than the free stream value. The reason for this is that for large $s_\mathrm{x}$ the wake recovery of the wake model is matched to the recovery predicted by the ``top-down'' model. The wake model is inherently less accurate in the fully developed regime when $s_\mathrm{x}$ is large. This inaccuracy is due in part to the following factors: (I) the wake expansion may not be linear in the fully developed region, (II) adding wake interactions using equation \eqref{eq-velsuperposed} could miss some effects, (III) the wake expansion in the vertical direction assumed in this model is not limited by the maximum internal boundary layer thickness. The expansion coefficient for the staggered $s_\mathrm{y}=3.49$ case shows a marked uptick for $s \sim 10$. Sideways wakes at some point reach the turbine of interest (reference turbine in the fully developed regime) which leads to an increase in the wake effects (up to the point that the reference turbine is fully in the spanwise wake) for a range of streamwise spacings. The CWBL model adjusts the $k_\mathrm{w}$ value to match with the ``top-down'' model and this can lead to a non-monotonic behavior of $k_\mathrm{w}$ as function of $s$ especially for staggered farms.

The panels (c) and (d) of figure \ref{figure10} show the effective spanwise spacing $s_\mathrm{ye}$ obtained with the CWBL model. For the aligned configuration we see that $s_\mathrm{ye}\approx 3.5$ for most cases. For this reason increasing the spacing beyond this value does not increase the power output in downstream turbine rows for the aligned configuration. Figure \ref{figure9}a shows that this predicted trend is in agreement with the LES data. 

\subsection{The entrance region of the wind-farms} \label{Section_Results_finite}

\begin{figure}
\centering
\includegraphics[width=0.99\textwidth]{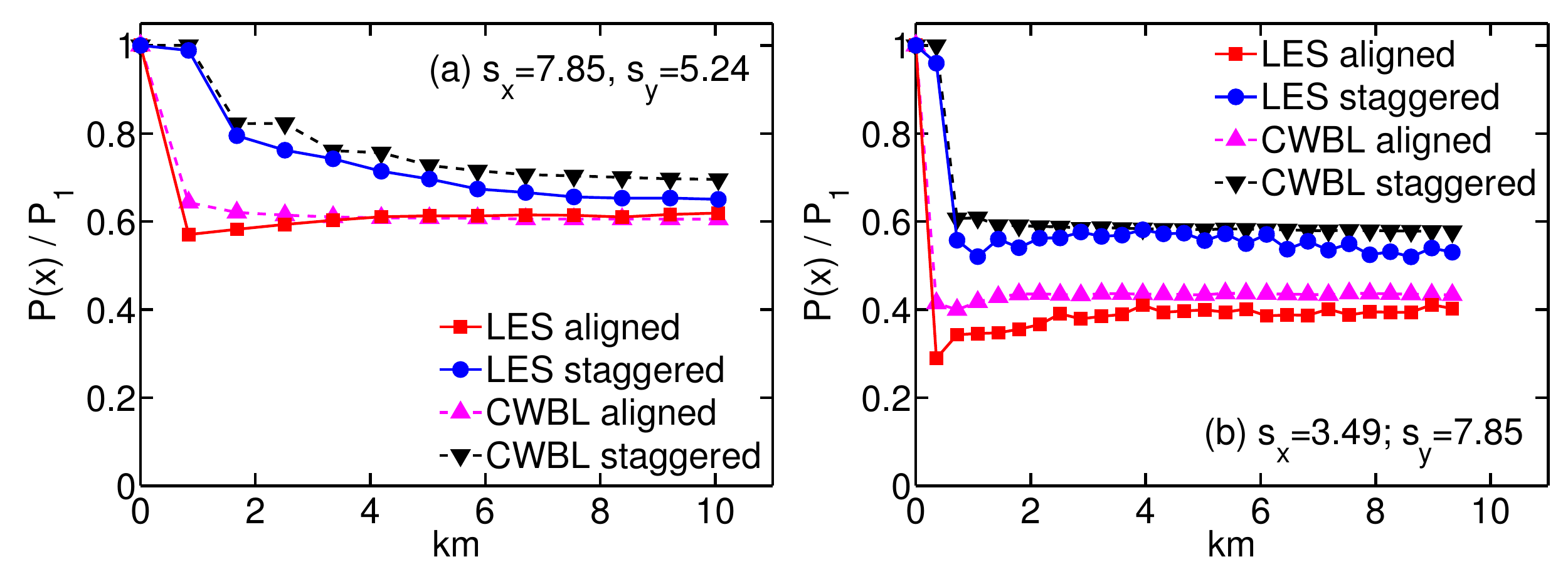}
\caption{Comparison of the power development as function of downstream distance obtained from LES \cite{ste13,ste14b,ste14f} and the CWBL model. Panels (a) and (b) indicate the results for two different combinations of the streamwise and spanwise spacings as indicated.}
\label{figure11}
\end{figure}

\begin{figure}
\centering
\includegraphics[width=0.99\textwidth]{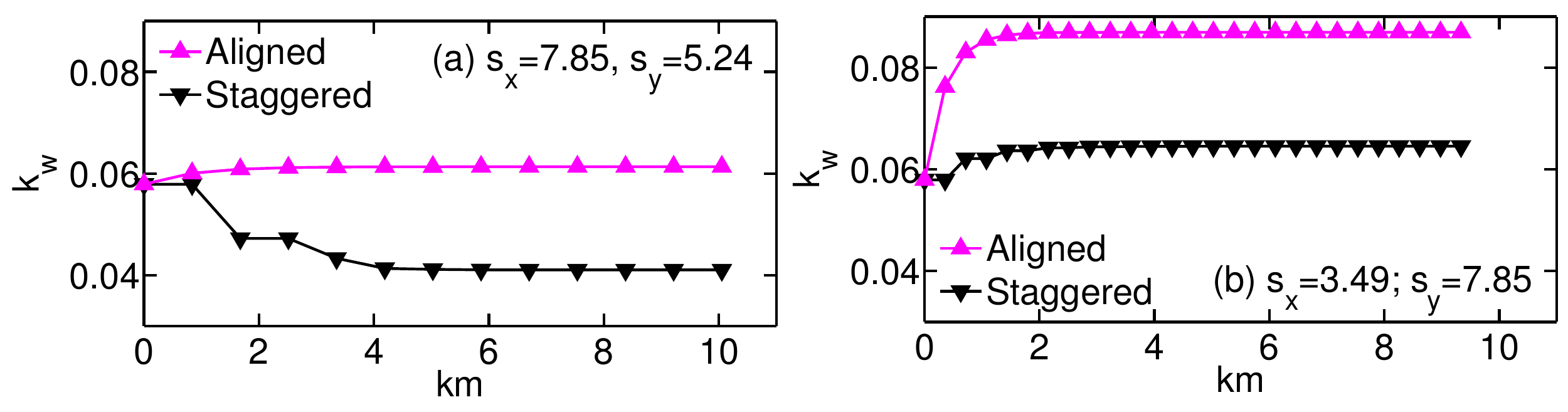}
\caption{Development of the wake expansion coefficient for the cases shown in figure \ref{figure11}. Panels (a) and (b) indicate the results for two different combinations of spanwise and streamwise spacing as indicated. }
\label{figure12}
\end{figure}

\begin{figure}
\centering
\includegraphics[width=0.99\textwidth]{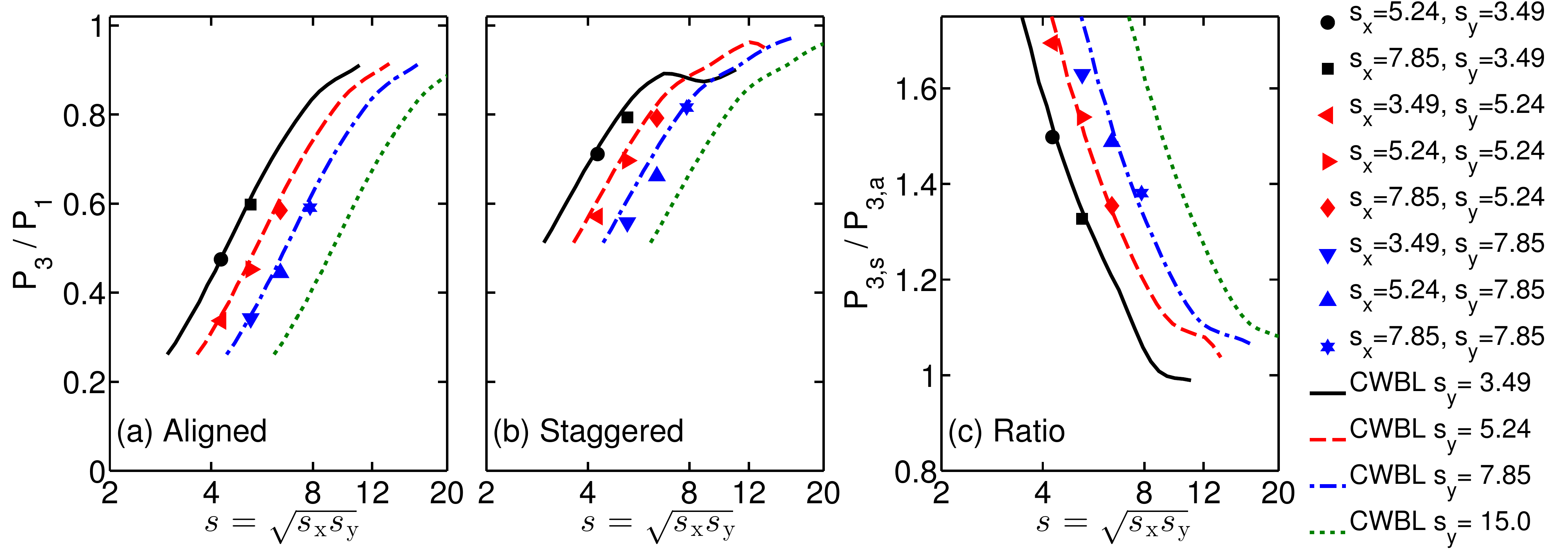}
\caption{Comparison of the CWBL model (lines) and LES results (symbols) \cite{ste13,ste14b,ste14f} for the relative power output for the third row ($P_\mathrm{3}/P_\mathrm{1}$) in (a) aligned and (b) staggered wind-farms as a function of the geometric mean turbine spacing $s=\sqrt{s_\mathrm{x}s_\mathrm{y}}$. Panel (c) gives the ratio between the staggered and aligned case $P_\mathrm{3,s}/P_\mathrm{3,a}$.}
\label{figure13}
\end{figure}

In this section the results of the CWBL model for the entrance region of the wind-farm are compared with LES. Figure \ref{figure11} shows the downstream power development for aligned and staggered wind-farms with different combinations of the spanwise and streamwise turbine spacings. From the figure we can see that the power output as function of the streamwise distance is captured well by the model. The differences observed for the fully developed state are in agreement with the differences seen in section \ref{Section_Results_fully}. Figure \ref{figure12} shows the development of the wake expansion coefficient for the cases shown in figure \ref{figure11}. This figure shows that the main changes in the wake expansion coefficient occur in the beginning of the wind-farm as given by equation \eqref{equation_finitewind-farm}.

Figure \ref{figure13} compares the relative power output at the third row predicted by the CWBL model with LES results. Again we see the model predicts the trends in the observed data well for aligned and staggered configurations. Comparing the results with the results for the fully developed region reveals that the benefit of the staggered over the aligned configuration is larger at the entrance of the wind-farm than in the fully developed state of the wind-farm. This observation is consistent with expectations and the results obtained from LES. A close look at figure \ref{figure13}b reveals a small decrease of the power output with increasing streamwise distance when $s_\mathrm{y}=3.49$ and the streamwise distance $s_\mathrm{x}\gtrsim20$. This is a feature of the regular wake model and is a result of spanwise wake expansion that affects the turbine of interest. 

\subsection{Comparisons of entire hub-height velocity field} \label{Section_Details}

\begin{figure}
\centering
\includegraphics[width=0.70\textwidth]{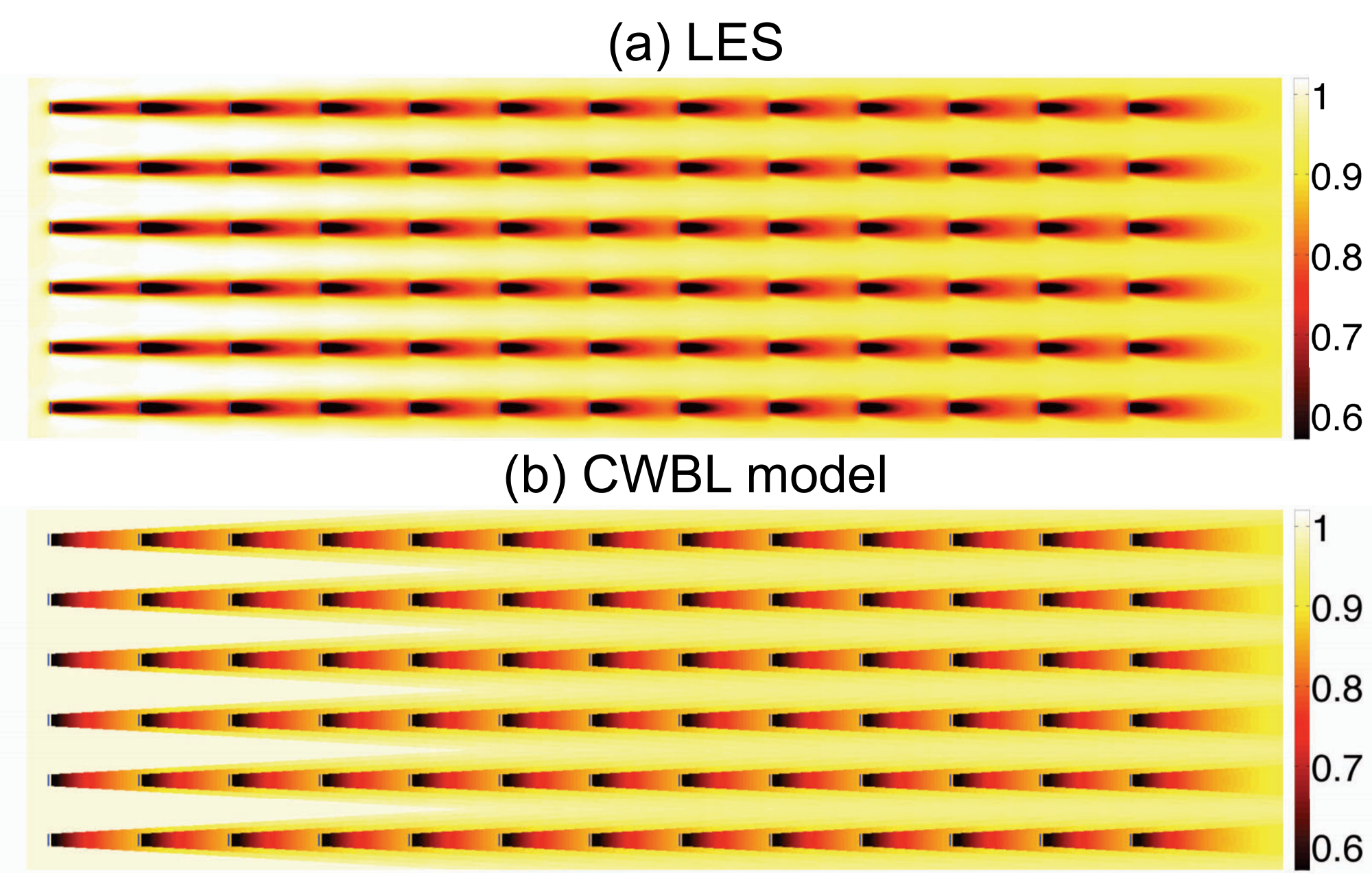}
\caption{Comparison between the (a) LES and (b) model averaged normalized hub-height velocity in an aligned wind-farm with a streamwise spacing of $s_\mathrm{x}=7.85$ and a spanwise spacing $s_\mathrm{y}=5.24$.}
\label{figure14}
\end{figure}

\begin{figure}
\centering
\includegraphics[width=0.70\textwidth]{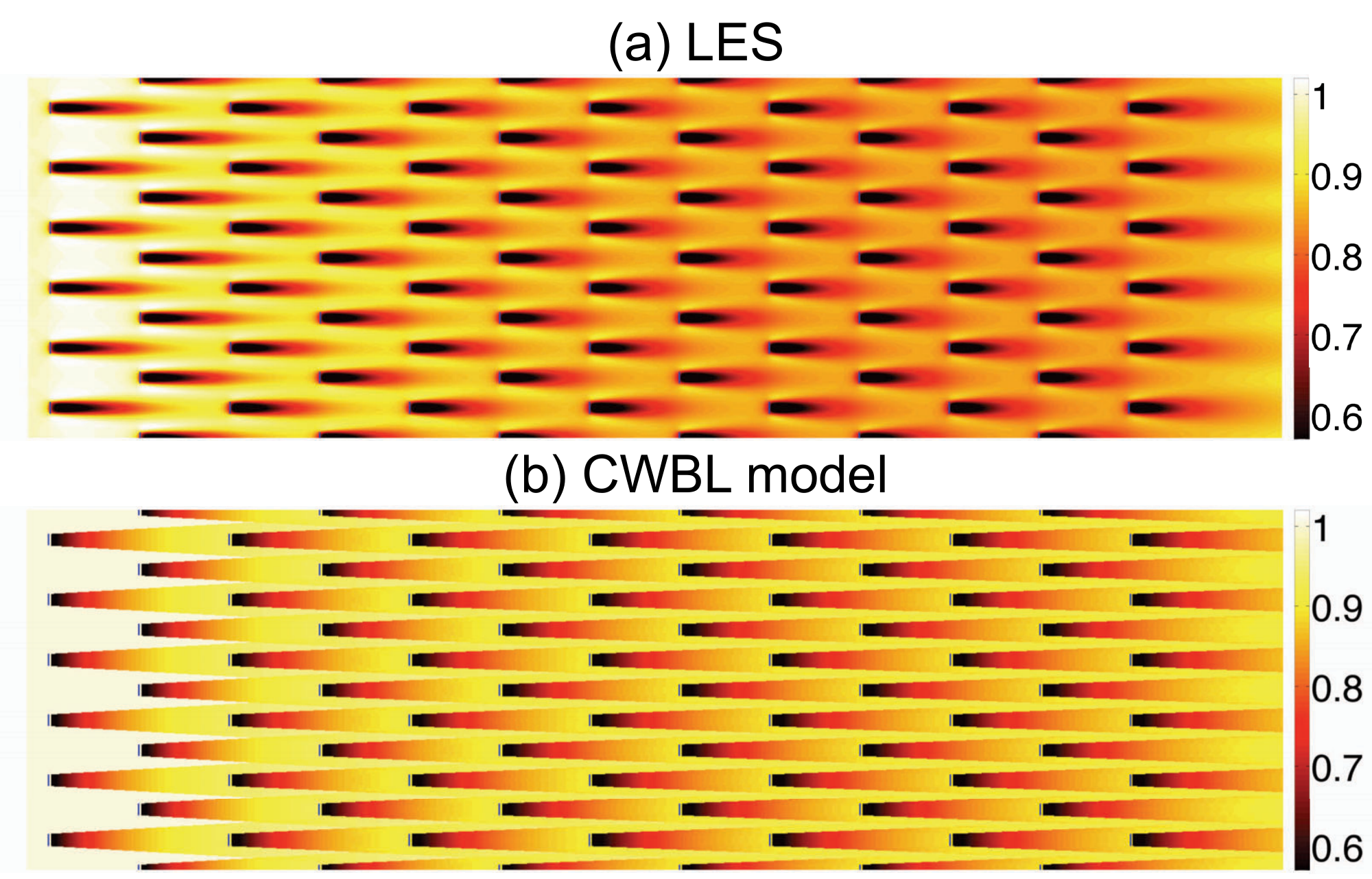}
\caption{Comparison between the (a) LES and (b) model averaged normalized hub-height velocity in a staggered wind-farm with a streamwise spacing of $s_\mathrm{x}=7.85$ and a spanwise spacing $s_\mathrm{y}=5.24$.}
\label{figure15}
\end{figure}

\begin{figure}
\centering
\includegraphics[width=0.70\textwidth]{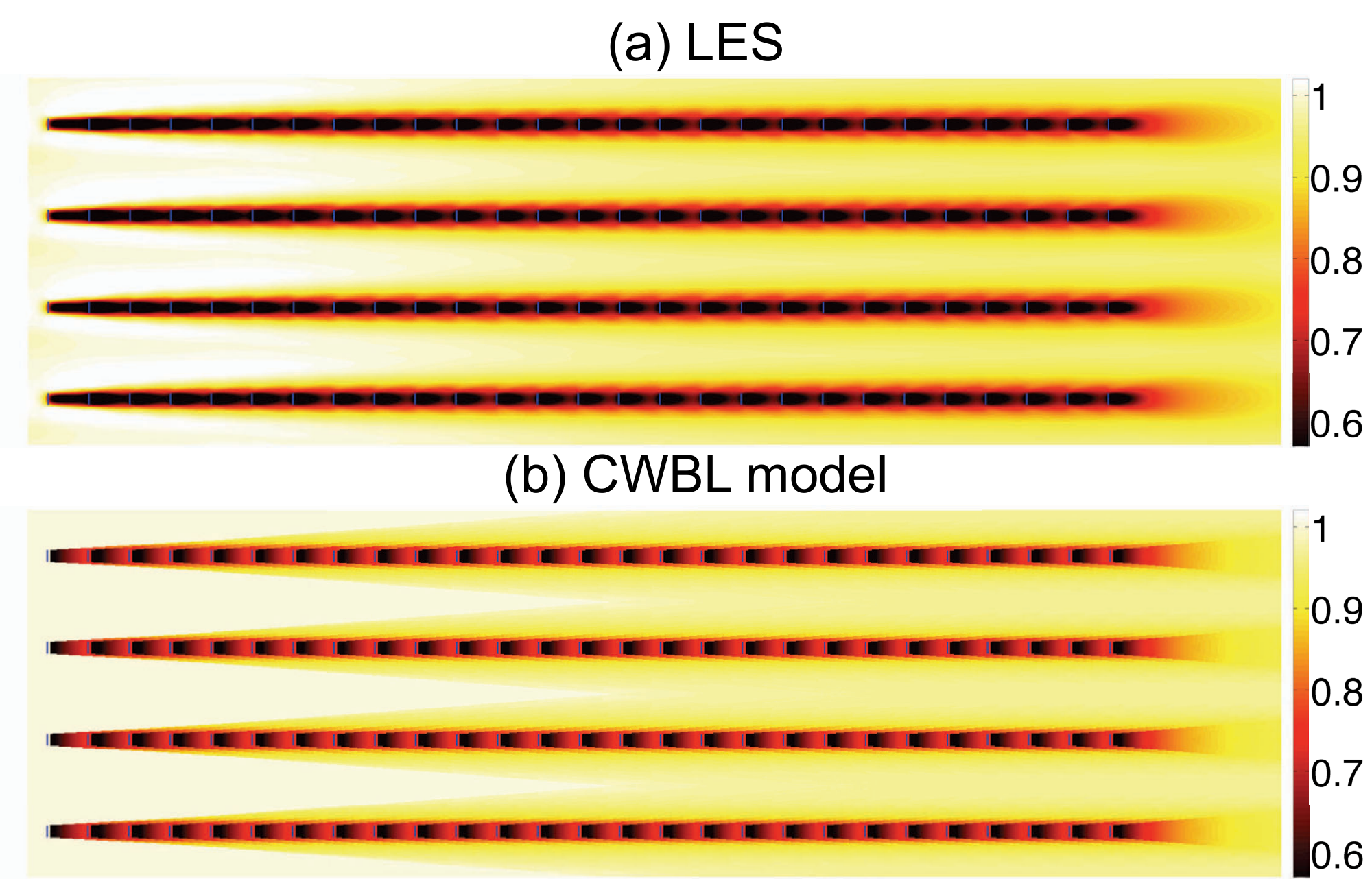}
\caption{Comparison between the (a) LES and (b) model averaged normalized hub-height velocity in an aligned wind-farm with a streamwise spacing of $s_\mathrm{x}=3.49$ and a spanwise spacing $s_\mathrm{y}=7.85$.}
\label{figure16}
\end{figure}

\begin{figure}
\centering
\includegraphics[width=0.70\textwidth]{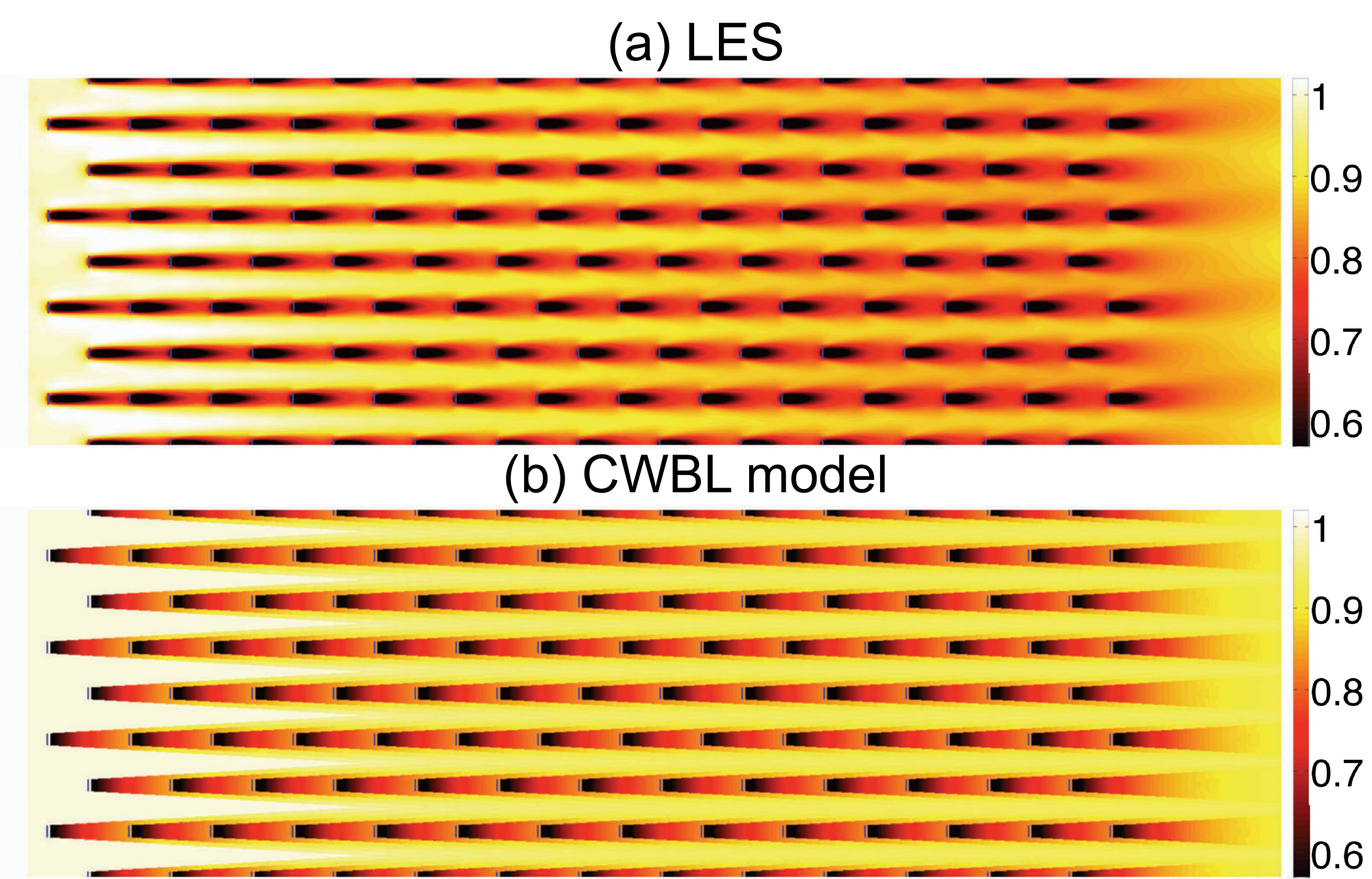}
\caption{Comparison between (a) LES and (b) model averaged normalized hub-height velocity in a staggered wind-farm with a streamwise spacing of $s_\mathrm{x}=3.49$ and a spanwise spacing $s_\mathrm{y}=7.85$.}
\label{figure17}
\end{figure}

\begin{figure}
\centering
\includegraphics[width=0.80\textwidth]{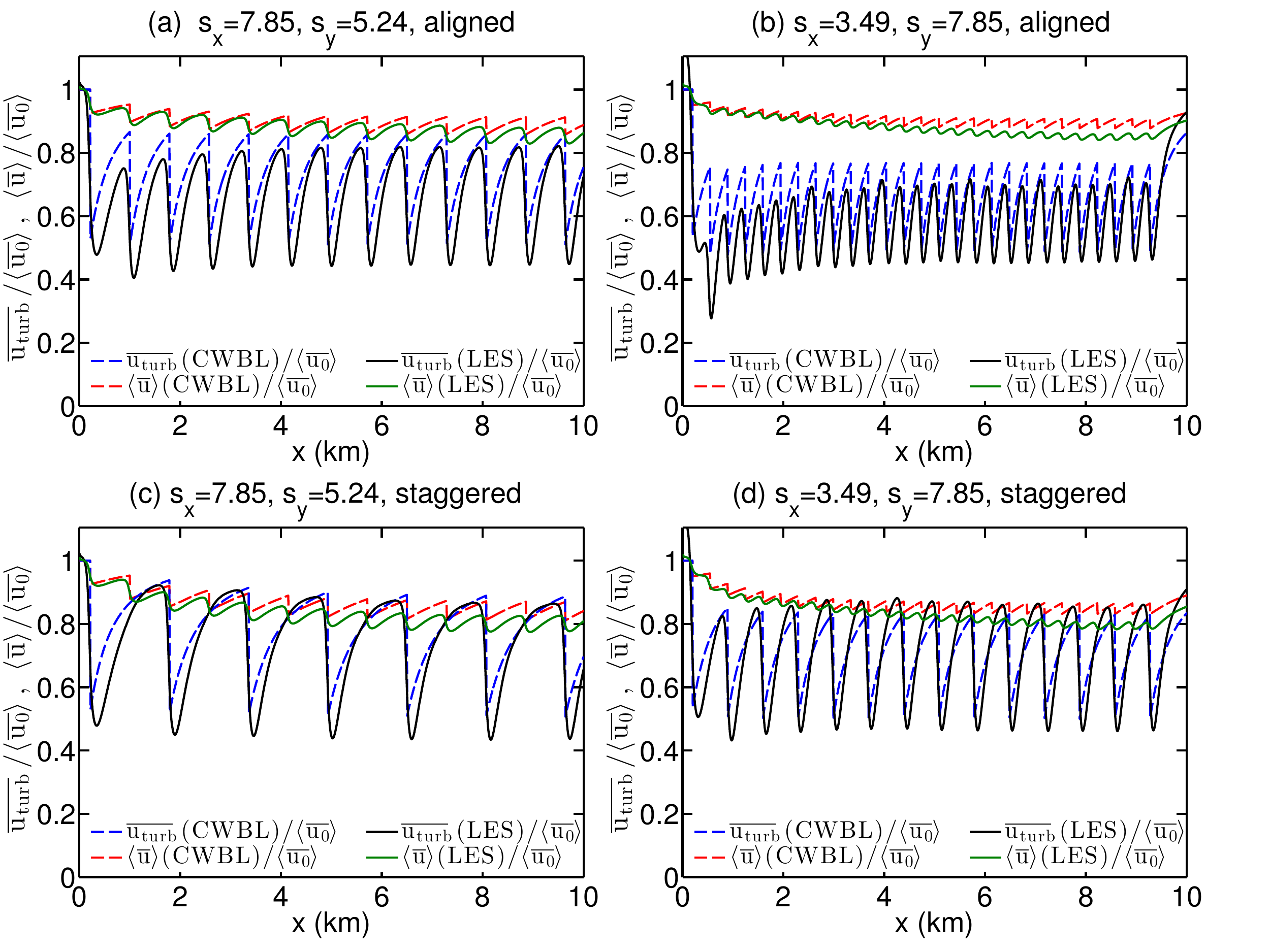}
\caption{A comparison of the horizontally and time averaged mean $\langle \overline{u} \rangle$ and velocity in the turbine row $\overline{u_\mathrm{turb}}$ as function of the downstream position for four different cases obtained from the LES and the CWBL model. The highest velocities are obtained just before the streamwise location of the consecutive turbine rows. Note that for a staggered wind-farm only every second turbine row has a turbine at a given spanwise location. Therefore the mean velocity $\langle \overline{u} \rangle$ has twice as many peaks for the staggered configuration as the velocity in a given turbine row $\overline{u_\mathrm{turb}}$. The results are normalized with the incoming velocity at hub-height $\langle \overline u_0 \rangle$.}
\label{figure18}
\end{figure}

Both the CWBL model and LES allow one to study the downstream development of velocities in the entire wind-farm. Figures \ref{figure14} to figure \ref{figure17} compare the velocity at hub-height obtained from the model with the LES for different cases. In agreement with what we have seen before we see that the CWBL model captures the main features of the LES. However, there are certain differences such as the exact wake recovery rate as function of the downstream distance and the precise way the velocity deficits progress further inside the farm. These effects can be made more quantitative by extracting the mean velocity at hub-height and the velocity in one of the turbine rows as function of the downstream position. Figure \ref{figure18} shows that the recovery of the wind velocity in the turbine rows is somewhat different in LES than in the model. We believe this is an effect of the wake-wake interactions that are not fully captured in the CWBL framework. As a consequence the horizontally averaged mean velocity at hub-height predicted by the model is not always accurate.

\subsection{Comparison with Horns Rev and Nysted data} \label{Section_HornsRev}

\begin{figure}
\centering
\includegraphics[width=0.80\textwidth]{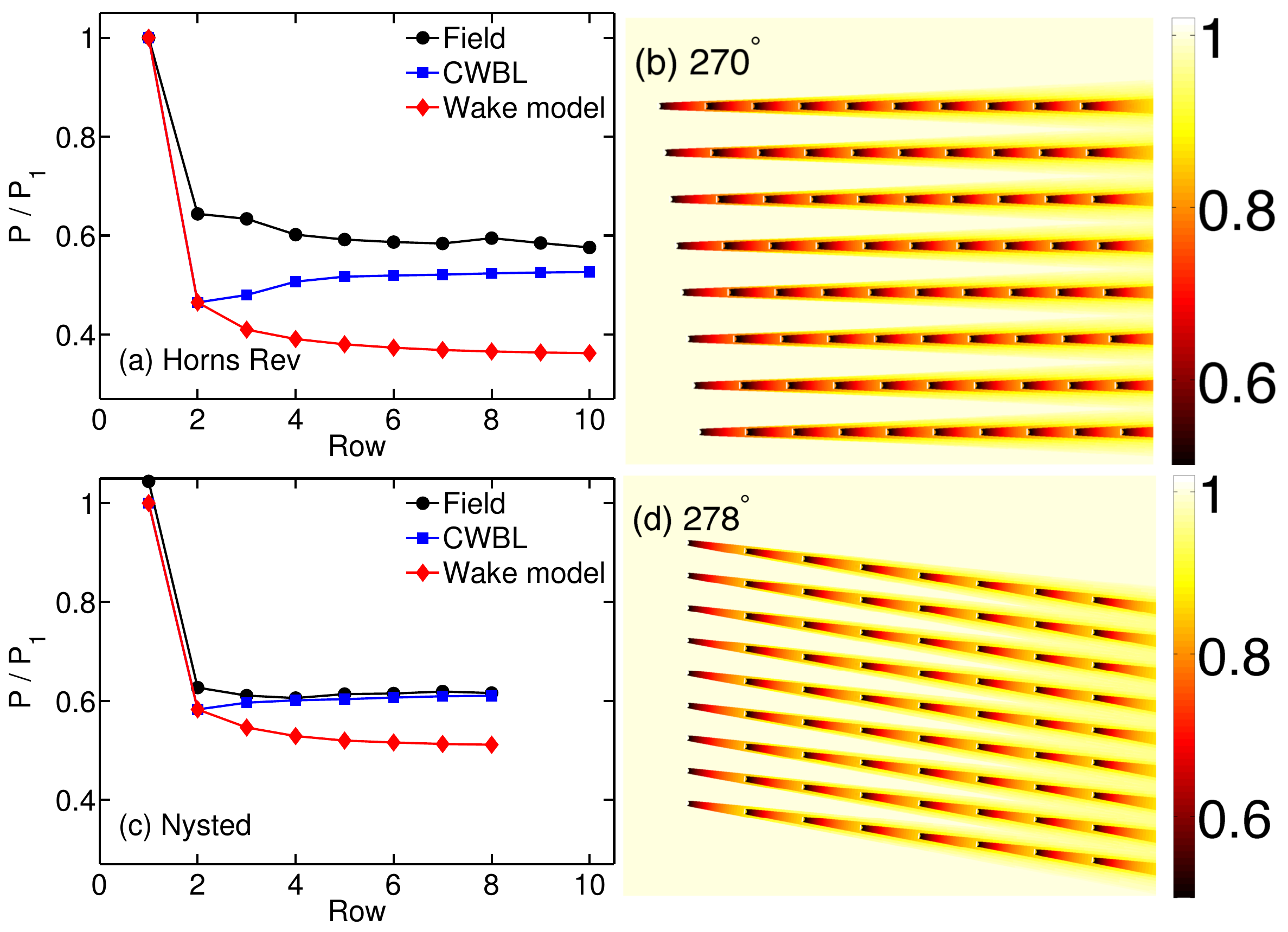}
\caption{Power degradation (a,c) and normalized hub-height velocities (b,d) in the Horns Rev and Nysted wind-farms for a wind-speed of $8 \pm 0.5$m/s and averaged over a 5 degree sector around the symmetry axis of the two wind-farms. The field data (from Ref. \cite{bar09c}, their figure 2) are shown as circles while the prediction from the CWBL model is shown by the squares. The top panels (a,b) indicate the results for Horns Rev (270$^\circ$) and the lower panels (c,d) for Nysted (278$^\circ$).}
\label{figure19}
\end{figure}

In this section we briefly illustrate how the model can be applied to an operational wind-farm using two well-known test cases, i.e.\ the aligned configuration for the Horns Rev and Nysted wind-farms. We apply the CWBL model to these wind-farms and compare the power degradation data for aligned flow from Ref. \cite{bar09c}. Specifically, for Horns Rev we use $s_\mathrm{x}=7.00$, $s_\mathrm{y} = 6.95$ as the layout parameters for the aligned flow configuration (270$^\circ$) and $s_\mathrm{x}=10.4$ and $s_\mathrm{y}=5.8$ for the aligned configuration of Nysted at 278$^\circ$. Horns Rev consists of Vestas V-80 2 MW turbines each with a hub-height of $z_{\mathrm{h}}=70$m and a rotor diameter $D=80$m. The turbines at Nysted have the parameters $z_{\mathrm{h}}=69$m and $D=82.4$m. As the wind-speed for the data we compare to is $8\pm0.5$ m/s we use $C_\mathrm{T} = 0.78$ \cite{por13,bar10b}. The height of the internal boundary layer is set to $500$ meters, i.e.\ the value used in the LES of Horns Rev by Port\'e-Agel {\it et al.}\ \cite{por13}. The surface roughness length $z_\mathrm{0,lo}=0.002$m is chosen to match the turbulence intensity of $7.7\%$ used in the Horns Rev LES by Port\'e-Agel {\it et al.}\ \cite{por13} at hub-height assuming logarithmic laws for the mean $\langle \overline{u} \rangle / u_*= \kappa^{-1} \log (z/z_\mathrm{0,lo})$ and variance $\langle \overline{(u^{\prime+})^2} \rangle = B_1 - A_1 \log (z/\delta)$ in the boundary layer with $A_1\approx1.25$ and $B_1 \approx 1.60$ \cite{mar13,men13,ste14d}. This results in a wake coefficient $k_\mathrm{w}=0.0382$ that is used at the entrance of the wind-farm in the CWBL model calculations, see section \ref{Section_combined_b}, and for the wake model results that are shown for comparison. The predicted power degradation with streamwise distance is shown in figure \ref{figure19}. Figure \ref{figure19} shows reasonably good agreement between the CWBL model and the field data. These results are promising but further work needs to be done such as contrasting these predictions with those obtained using other models as summarized in Ref.\ \cite{bar09c,san11,ste13,mor14,nyg14}. More cases and further tests, including a comparison with the LES study of Horns Rev provided by Port\'e-Agel, Wu and Chen \cite{por13}, will be considered elsewhere \cite{ste15} and are not included here for sake of brevity. 

\section{Conclusions} \label{Section_Conclusions}
In this paper we have introduced the CWBL model, a framework for predicting the power output in both the entrance and fully developed regions of wind-farms. The method combines two well-known approaches, the wake model and the ``top-down'' boundary layer model thus resulting in the proposed coupled model. Both of the constitutive approaches have one parameter that needs to be determined. For the wake model this is the wake expansion coefficient $k_\mathrm{w}$ and for the ``top-down'' model this is the effective spanwise spacing $s_\mathrm{ye}$. In the CWBL model, the effective spanwise spacing is obtained from the wake model and is then used in the ``top-down'' model. These results are then coupled through an iterative procedure to obtain the wake expansion coefficient $k_\mathrm{w}$ that ensures that the turbine velocity is matched in both models. A detailed comparison with LES results for a variety of cases reveals that the model represents the LES data quite well for both the fully developed region and the entrance region of the wind-farm. The final part of the work illustrates application of the CWBL model to field-scale wind-farm data by comparing the power degradation measurements for the Horns Rev and Nysted wind-farms to those estimated using the CWBL model. Good agreement has been obtained. By combining relevant wake growth and boundary layer physics, the coupled model is promising and can be explored in further tests and applications \cite{ste15}. \\

{\bf Acknowledgements.} The authors thank Claire VerHulst for comments. This work is funded in part by the research program `Fellowships for Young Energy Scientists' (YES!) of the Foundation for Fundamental Research on Matter (FOM) supported by the Netherlands Organization for Scientific Research (NWO), and by the National Science Foundation grant IIA $1243482$ (the WINDINSPIRE project). This work used the Extreme Science and Engineering Discovery Environment (XSEDE), which is supported by the National Science Foundation grant number OCI-1053575 and the LISA and Cartesius clusters of SURFsara in the Netherlands.

\FloatBarrier

\section*{Appendix 1: Computationally efficient methods for wake model}

For computational reasons it can be convenient to have an approximation of the results obtained by the wake model in the fully developed regime. It has been shown by Pe$\tilde{\mbox{n}}$a and Rathmann \cite{pen14b} that such an approximation can be obtained for an aligned wind-farm configuration while assuming that a turbine experiences the full wake effects when the wake has reached the turbine center. This appendix present a generalization of this approach.
 
Following Pe$\tilde{\mbox{n}}$a and Rathmann \cite{pen14b} we define that the total wake deficit $\delta_T$ is given by
\begin{equation}
\label{appendix1-1}
\delta_T^2= \delta_{\mathrm{I}}^2+\delta_{\mathrm{II}}^2
\end{equation}
where $\delta_{\mathrm{I}}$ indicate the contributions from the turbines directly upstream (the turbines above ground as well as the 'ghost' turbines, i.e.\ turbines $J_U$) and $\delta_{\mathrm{II}}$ the contributions from adjacent turbines (again for the turbines above and below the ground, i.e.\ turbines $J_S$). The initial wake deficit $\delta_\mathrm{0}$ is given by
\begin{equation}
\label{appendix1-2}
\delta_\mathrm{0}=2a=1-\sqrt{1-C_\mathrm{T}}.
\end{equation}
The total wake contributions for the aligned and staggered case can then be approximated by determining $\delta_{\mathrm{I}}^2$ and $\delta_{\mathrm{II}}^2$ as indicated below. 

Parenthetically, we note that it is assumed implicitly that the initial area of the wake corresponds to the turbine disk area rather than the slightly enlarged area appropriate for the velocity reduction $2a$. If the wake is assumed to begin at the enlarged stream tube area behind the turbine, the denominator $(1+k_\mathrm{w} x /R)^2$ should be replaced by $(\gamma+k_\mathrm{w} x /R)^2$ where $\gamma = [(1-a)/(1-2a)]^{1/2}$. For typical values of $a$, $\gamma$ can in fact be quite a bit larger than 1. The usual wake model \cite{jen83} assumes instead that the wake with a deficit $2a ~u_{0}$ begins at the smaller turbine area $\pi R^2$. Here we have followed the same standard approach but keep in mind that future improvements may be required to make the entire approach more internally self-consistent.

\subsection*{Aligned configuration} \label{Section_sumsaligned}

\begin{figure}
\centering
\includegraphics[width=0.99\textwidth]{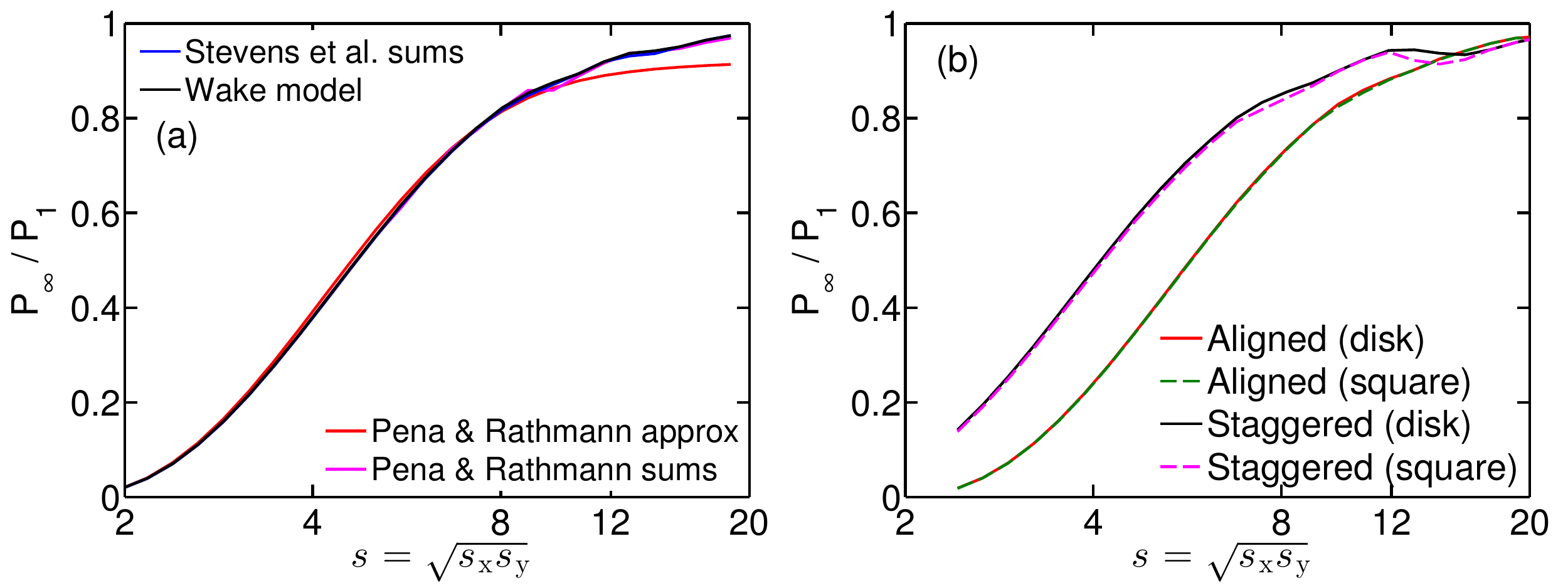}
\caption{Power output in the fully developed regime according to the wake model with $k=0.0579$. a) Comparison between the wake model calculations and different approximations. b) Comparison of rectangular and circular turbines and wakes.}
\label{figure20}
\end{figure}
For an aligned infinite wind-farm $\delta_{\mathrm{I}}^2$ can be approximated as
\begin{eqnarray}
\label{appendix1-3}
&&\delta_{\mathrm{I}}^2 = \delta_\mathrm{0}^2 \sum_{j=1}^\infty (1+c_\mathrm{w}) \left(1+2 k_\mathrm{w} s_\mathrm{x} j \right)^{-4} \mbox{~~~~with}\\
&& a_\mathrm{w}=(k_\mathrm{w} s_\mathrm{x} j)-(2z_{\mathrm{h}}/D-1), \mbox{~~~~} b_\mathrm{w}=\min(a_\mathrm{w},1), \mbox{~~~~} c_\mathrm{w}=\max(b_\mathrm{w},0). \nonumber
\end{eqnarray}
Here the term $\left(1+2 k_\mathrm{w} s_\mathrm{x} j \right)^{-4}$ indicates the squared velocity deficit resulting from an upstream turbine. The $(1+c_\mathrm{w})$ term indicates whether the wakes have reached the turbine of interest. The turbine of interest will always be completely in the wake of directly upstream turbines (which is represented by the $1$). The $c_\mathrm{w}$ estimates the fraction of the turbine of interest that is covered by wakes originating from the directly upstream 'ghost' turbines and is defined such that it is between $0$ and $1$. 

The wake contributions for the adjacent and adjacent 'ghost' turbines, i.e.\ $\delta_{\mathrm{II}}^2$, are approximated as
\begin{eqnarray}
\label{appendix1-4}
&&\delta_{\mathrm{II}}^2 = \delta_\mathrm{0}^2 \sum_{\mathrm{i}=1}^{\infty} \sum_{j=1}^\infty 2(f_\mathrm{w}+g_\mathrm{w}) \left(1+2 k_\mathrm{w} s_\mathrm{x} j \right)^{-4} \mbox{~~~~with}\\
&& a_\mathrm{w}=(k_\mathrm{w} s_\mathrm{x} j)-(2z_{\mathrm{h}}/D-1), \mbox{~~~~} b_\mathrm{w}=\min(a_\mathrm{w},1), \mbox{~~~~} c_\mathrm{w}=\max(b_\mathrm{w},0) \nonumber \\
&& d_\mathrm{w}=(k_\mathrm{w} s_\mathrm{x} j)-(s_\mathrm{y}-1+(\mbox{i}-1)s_\mathrm{y}), \mbox{~~~~} e_\mathrm{w}=\min(d_\mathrm{w},1), \mbox{~~~~} f_\mathrm{w}=\max(e_\mathrm{w},0) \nonumber \\
&& g_\mathrm{w}=c_\mathrm{w}f_\mathrm{w}. \nonumber
\end{eqnarray}
Just as above $c_\mathrm{w}$ gives the fraction of the turbine that is covered by wakes created from upstream 'ghost' turbines, while $f_\mathrm{w}$ determines the fraction of the turbine that is covered by wakes created from adjacent turbines. The factor $g_\mathrm{w}$ determines the fraction of the turbine that is covered by wakes created from the adjacent `ghost' turbines. The factor $2(f_\mathrm{w}+g_\mathrm{w})$ adds the effects of the adjacent and adjacent 'ghost' turbine rows on the left and right side of the turbine of interest. We note that figure \ref{figure20}a show that this is a very good approximation for the aligned configuration. 

\subsection*{Staggered configuration} \label{Section_sumsstagger}
For a staggered wind-farm $\delta_{\mathrm{I}}^2$ and $\delta_{\mathrm{II}}^2$ can be approximated in a similar way as for the aligned case. For $\delta_{\mathrm{I}}^2$ it becomes
\begin{eqnarray}
\label{appendix1-5}
&&\delta_{\mathrm{I}}^2 = \delta_\mathrm{0}^2 \sum_{j=1}^\infty (1+c_\mathrm{w}) \left(1+2 k_\mathrm{w} s_\mathrm{x} (2j) \right)^{-4} \mbox{~~~~with}\\
&& a_\mathrm{w}=(k_\mathrm{w} s_\mathrm{x} (2j))-(2z_{\mathrm{h}}/D-1), \mbox{~~~~} b_\mathrm{w}=\min(a_\mathrm{w},1), \mbox{~~~~} c_\mathrm{w}=\max(b_\mathrm{w},0). \nonumber
\end{eqnarray}
Here the term $2j$ makes sure that we have a staggered configuration (direct upstream turbines every other row). Similarly, $\delta_{\mathrm{II}}^2$ is approximated as 
\begin{eqnarray}
\label{appendix1-6}
&&\delta_{\mathrm{II}}^2 = \delta_\mathrm{0}^2 \sum_{\mathrm{i}=1}^{\infty}\sum_{j=1}^\infty 2(f_\mathrm{w}+g_\mathrm{w}) \left(1+2 k_\mathrm{w} s_\mathrm{x} (2j-1) \right)^{-4} \mbox{~~~~with}\\
&& a_\mathrm{w}=(k_\mathrm{w} s_\mathrm{x} (2j-1))-(2z_{\mathrm{h}}/D-1), \mbox{~~~~} b_\mathrm{w}=\min(a_\mathrm{w},1), \mbox{~~~~} c_\mathrm{w}=\max(b_\mathrm{w},0) \nonumber \\
&& d_\mathrm{w}=(k_\mathrm{w} s_\mathrm{x} (2j-1))-(s_\mathrm{y}-1+(\mbox{i}-1)s_\mathrm{y}), \mbox{~~~~} e_\mathrm{w}=\min(d_\mathrm{w},1), \mbox{~~~~} f_\mathrm{w}=\max(e_\mathrm{w},0) \nonumber \\
&& g_\mathrm{w}=c_\mathrm{w}f_\mathrm{w} \nonumber
\end{eqnarray}
where the term $2j-1$ selects that we only have adjacent and adjacent `ghost' turbines every other row.

\subsection*{Results} \label{Section_Results_Jensen}
In figure \ref{figure20} the results of the wake model are compared with the approximation given in this appendix and the results from Pe$\tilde{\mbox{n}}$a and Rathmann \cite{pen14b}. The figure reveals that our approximation reproduces the results from the wake model very well in the fully developed regime. A comparison with the sum approximation by Pe$\tilde{\mbox{n}}$a and Rathmann \cite{pen14b} reveals good agreement between the two methods although our approximation is smoother when partial wake overlaps are important. We note that in the above approximations it is assumed that the turbines and wakes are square, just as Pe$\tilde{\mbox{n}}$a and Rathmann \cite{pen14b}. Figure \ref{figure20}b shows this is a reasonable assumption as a comparison of both cases only shows small differences due to this approximation. The approximations given in this appendix can be useful for an efficient implementation of the wake model coupled with the ``top-down'' model.

\section*{Appendix 2: Additional details on ``top-down'' model}

As the ``top-down'' model uses horizontal averaging it only knowns one velocity scale. This implies that the model assumes that the velocity in front of the turbines $u_{\mathrm{turb}}$ should be equal to the horizontally averaged mean velocity $u_{\mathrm{mean}}$ when averaging over the appropriate spanwise $s_\mathrm{ye}$ region. It is not obvious that this condition is always met. From figure \ref{figure5}b we see that the ``top-down'' model predicts the power output of the staggered case very well, i.e.\ cases in which $s_\mathrm{y}^*$ is larger than the actual $s_\mathrm{y}$ such that it does not influence the ``top-down'' model calculations. This observation indicates that the ``top-down'' model predicts the velocity in front of the turbines very well. Below we show with results from LES that this observation is consistent with the measured mean velocity profiles from LES. We think the agreement stems from the use of the velocity scale in equation \eqref{Eq_momentum_balance} to calculate the momentum loss leading to predictions of the mean velocity profile closer to the turbine velocity than to the mean velocity.

\begin{figure}
\includegraphics[width=0.99\textwidth]{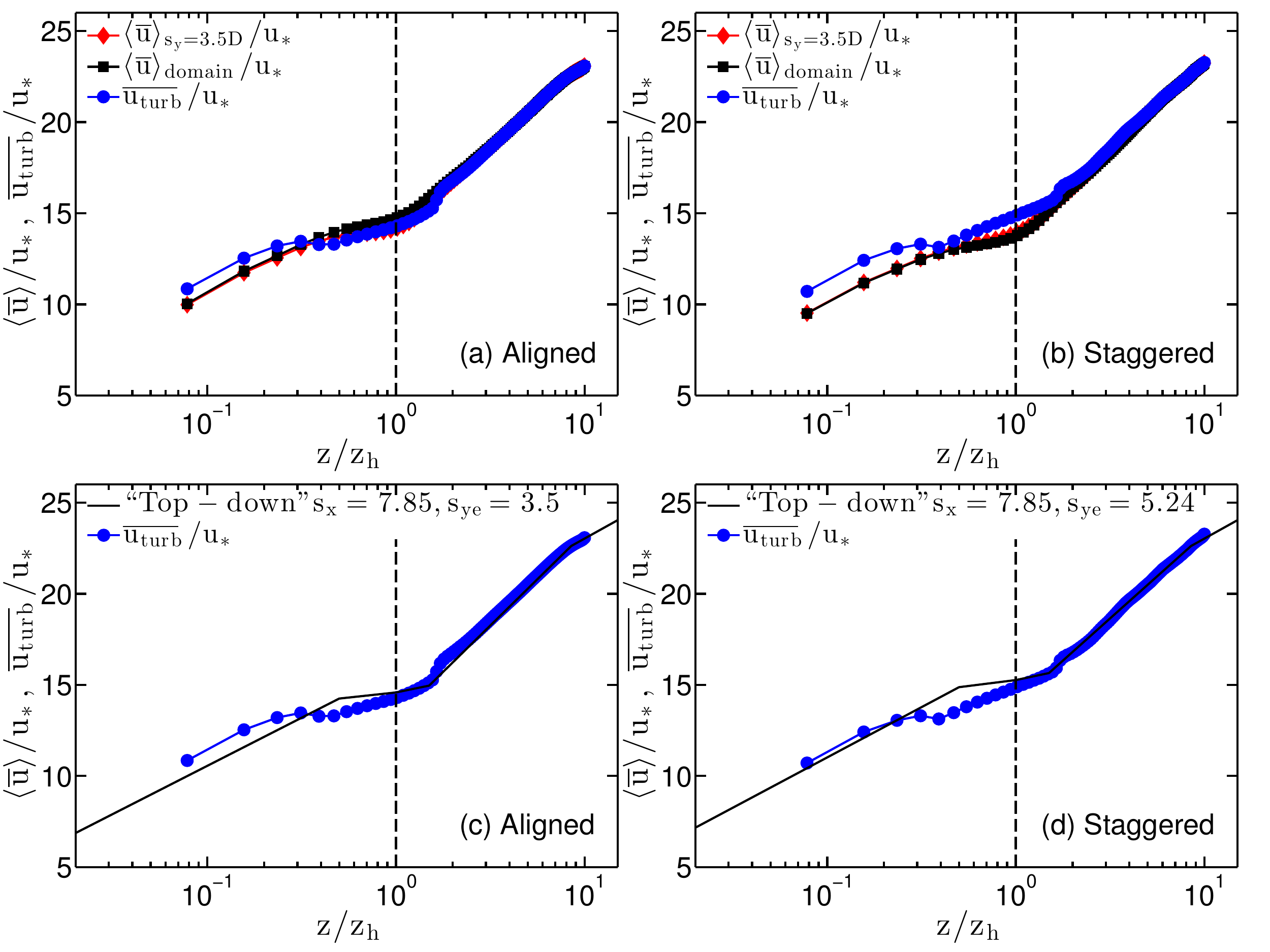}
\caption{Comparison of the vertical velocity profile obtained using the ``top-down'' model with velocities obtained from an infinitely large staggered wind-farm with $s_\mathrm{x}=7.85$ and $s_\mathrm{y}=5.24$. Panels (a) and (b) show the vertical profiles of the streamwise velocity averaged over the complete horizontal ($s_\mathrm{y}=5.24$, squares), averaged over a $s_\mathrm{y}=3.5$ region around the turbines (diamonds), and turbine velocity (circles). Panels (c) and (d) compare the turbine velocity obtained from the LES data with the ``top-down'' model predictions with the appropriate $s_\mathrm{ye}$.}
\label{figure21}
\end{figure}

The results in figure \ref{figure21} are from simulations of infinitely large wind-farms \cite{ste14e}, as the available symmetries there allow for more averaging and therefore better comparisons then the developing cases. For the cases here the streamwise spacing $s_\mathrm{x}=7.85$ and the spanwise spacing is $s_\mathrm{y}=5.24$. The results in figure \ref{figure21} show $\overline{u_{\mathrm{turb}}}$, $\overline{u_{\mathrm{mean}}}$, and $\overline{u_{\mathrm{mean}}}$ averaged using a smaller spanwise distance of $3.5D$ centered around the turbines. For the aligned case the smaller spanwise area is roughly equal to $s_\mathrm{ye}$ and over this region $u_{\mathrm{turb}}$ and the local mean are almost the same. As a result the predicted velocity by the ``top-down'' model agrees at hub-height with both velocities. For the staggered case the situation is more complicated. Here $s_\mathrm{y}^*$ is larger than the actual spanwise spacing, so the relevant averaging interval should be the whole horizontal area. However, the figure shows that using this interval $\overline{u_{\mathrm{turb}}}$ and $\overline{u_{\mathrm{mean}}}$ are not the same. A comparison with the predicted ``top-down'' velocities shows its prediction is much closer to $\overline{u_{\mathrm{turb}}}$ than to $\overline{u_{\mathrm{mean}}}$ as argued above. 

\end{document}